\documentclass[12pt,aps,prb]{revtex4}
\usepackage{epsfig}
\usepackage{amsmath}

\normalsize

\def\bk{{\bf k}}
\def\bn{{\bf n}}
\def\bp{{\bf p}}
\def\bq{{\bf q}}

\def\bt{{\bf t}}
\def\bv{{\bf v}}

\def\b0{{\bf 0}}

\def\cO{{\cal O}}

\def\Re{{\rm Re}}
\def\Im{{\rm Im}}
\def\bra{\langle}
\def\ket{\rangle}

\def\alf{\alpha}
\def\eps{\epsilon}
\def\gam{\gamma}
\def\Gam{\Gamma}

\def\Lam{\Lambda}
\def\om{\omega}

\def\sg{\sigma}
\def\Sg{\Sigma}

\def\sgn{{\rm sgn}}

\begin{document}


\title{Fermi surface fluctuations and single electron excitations \\
 near Pomeranchuk instability in two dimensions}

\author{Luca Dell'Anna and Walter Metzner}

\affiliation{Max-Planck-Institut f\"ur Festk\"orperforschung, 
 D-70569 Stuttgart, Germany}

\date{\small\today}


\begin{abstract}

A metallic electron system near an orientational symmetry
breaking Pomeranchuk instability is characterized by a ''soft''
Fermi surface with enhanced collective fluctuations. 
We analyze fluctuation effects in a two-dimensional electron
system on a square lattice in the vicinity of a Pomeranchuk 
instability with d-wave symmetry, using a phenomenological model 
which includes interactions with a small momentum transfer only.
We compute the dynamical density correlations with a d-wave 
form factor for small momenta and frequencies, the dynamical
effective interaction due to fluctuation exchange, and the 
electron self-energy. 
At the quantum critical point the density correlations and
the dynamical forward scattering interaction diverge with a
dynamical exponent $z=3$. 
The singular forward scattering leads to large self-energy
corrections, which destroy Fermi liquid behavior over the
whole Fermi surface except near the Brillouin zone diagonal.
The decay rate of single-particle excitations, which is related
to the width of the peaks in the spectral function, exceeds
the excitation energy in the low-energy limit. The dispersion 
of maxima in the spectra flattens strongly near those portions 
of the Fermi surface which are remote from the zone diagonal. \\
\noindent
\mbox{PACS: 71.10.Fd, 71.18.+y, 74.20.Mn}
\end{abstract}


\maketitle

\section{Introduction}

Under suitable circumstances electron-electron interactions can 
generate a spontaneous breaking of the rotation symmetry of an 
itinerant electron system without breaking translation invariance.
From a Fermi liquid perspective such an instability is driven by 
forward scattering interactions and leads to a symmetry breaking 
deformation of the Fermi surface. 
Alluding to a stability limit for forward scattering derived 
long ago by Pomeranchuk,\cite{Pom} it is therefore frequently
referred to as \emph{''Pomeranchuk instability''}.
On a square lattice, a Pomeranchuk instability with $d_{x^2-y^2}$ 
symmetry, where the Fermi surface expands along the $k_x$ axis 
and shrinks along the $k_y$ axis (or vice versa), was first 
considered for the t-J,\cite{YK1} Hubbard,\cite{HM,GKW} and 
extended Hubbard model.\cite{VV}
Fermi surface symmetry breaking in fully isotropic (not lattice)
systems has also been analyzed.\cite{OKF}
A Pomeranchuk instability leads to a state with the same symmetry
reduction as the ''nematic'' electron liquid defined by Kivelson
et al.\cite{KFE} in their discussion of similarities between doped 
Mott insulators with charge stripe correlations and liquid crystal 
phases.

A Pomeranchuk instability usually has to compete with other
instabilities, but can also coexist with other symmetry breaking
order. For example, in the two dimensional Hubbard model with a 
sizable next-to-nearest neighbor hopping and an electron density
near van Hove filling a superconducting state with a d-wave
deformed Fermi surface is stabilized at least at weak 
coupling.\cite{NM}
Superconducting nematic states have also been included in a
general classification of possible symmetry breaking patterns
by Vojta et al.\cite{VZS}
In this work we will however focus on symmetry breaking Fermi
surface deformations in an otherwise normal state.
Since a Pomeranchuk instability of electrons on a lattice 
breaks only a discrete symmetry, no Goldstone mode exists and
symmetry-broken states are stable also at finite temperature
in $d \geq 2$ dimensions.

Electron systems in the vicinity of a Pomeranchuk instability 
have peculiar properties due to a \emph{''soft'' Fermi surface}, 
which can be easily deformed by anisotropic perturbations.
In particular, it was shown that dynamical fluctuations of such 
a soft Fermi surface lead to a strongly enhanced decay rate for 
single-particle excitations and thus to non-Fermi liquid 
behavior.\cite{MRA} 
Close to a $d_{x^2-y^2}$ Pomeranchuk instability the decay rate 
is maximal near the $k_x$ and $k_y$ axes and minimal near the 
diagonal of the Brillouin zone. 
The decay rate was computed in random phase approximation (RPA)
for a phenomenological model describing electrons on a square 
lattice which interact exclusively via scattering processes with
small momentum transfers.\cite{MRA}
Subsequently it was shown that the putative continuous 
Pomeranchuk transition in this model is usually preempted by 
a first order transition at low temperatures, that is the
Fermi surface symmetry changes abruptly before the fluctuations
become truely critical.\cite{KKC,KCOK} 
However, for reasonable choices of hopping and interaction 
parameters the system is nevertheless characterized by a 
drastically softened Fermi surface on the symmetric side of
the first order transition, and hence by strongly enhanced
Fermi surface fluctuations.\cite{YOM}
Furthermore, adding a uniform repulsion to the forward 
scattering interaction, the first order transition can be 
suppressed, and for a favorable but not unphysical choice of
model parameters even a genuine quantum critical point can be 
realized.\cite{YOM}

In this work we present a detailed analysis of dynamical
Fermi surface fluctuations in the vicinity of a Pomeranchuk
instability with $d_{x^2-y^2}$-wave symmetry in two dimensions
and their effect on single electron excitations.
We compute the dynamical density correlations with a d-wave
form factor for small momenta and frequencies, the dynamical
effective interaction due to fluctuation exchange, and the
electron self-energy, from which the spectral function for
single-particle excitations is obtained.
In the quantum critical regime the density correlations and
the dynamical forward scattering interaction diverge with a
singularity familiar from other quantum phase transitions
in itinerant electron systems with dynamical exponent $z=3$.
The singular forward scattering leads to large self-energy
contributions $\Sg(\bk,\om)$, which are proportional to 
$d_{\bk}^2$, where $d_{\bk}$ is a form factor with d-wave 
symmetry.
At the quantum critical point, $\Im\Sg(\bk_F,\om)$ scales
to zero as $|\om|^{2/3}$.
Fermi liquid behavior is thus destroyed over the whole Fermi 
surface except near the Brillouin zone diagonal.
We also compute the crossover from non-Fermi to Fermi liquid 
behavior in case of a large finite correlation length $\xi$ 
in the ground state.
In the quantum critical regime at low finite temperatures
the self-energy has a part due to quantum fluctuations, which 
obeys $(\om/T)$-scaling, but also a classical part, which is
proportional to $T\xi(T)$ near the Fermi surface. The 
classical contribution violates $(\om/T)$-scaling and was
overlooked in Ref.\ \onlinecite{MRA}.

The article is structured as follows. In Sec.\ II we define our 
phenomenological model and review its mean-field phase diagram. 
In Sec.\ III we analyze the dynamical effective interaction of 
the model in a regime where the nearby Pomeranchuk instability
leads to strong Fermi surface fluctuations.
The low energy behavior of the self-energy in the presence of
these fluctuations is then computed in Sec.\ IV, and results
for the corresponding spectral function are discussed in 
Sec.\ V. We finally summarize and discuss the possible role 
of d-wave Fermi surface fluctuations in cuprate superconductors
in Sec.\ VI.

\section{f-model}

To analyze fluctuation effects in the vicinity of a Pomeranchuk
instability we consider a phenomenological lattice model with
an interaction which drives a Fermi surface symmetry breaking,
but no other instability. 
The model Hamiltonian, which has been introduced already in
Ref.\ \onlinecite{MRA}, reads
\begin{equation}
 H = \sum_{\bk,\sg} \eps_{\bk} \, n_{\bk\sg} +
 \frac{1}{2V} \sum_{\bk,\bk',\bq} f_{\bk\bk'}(\bq) \,
 n_{\bk}(\bq) \, n_{\bk'}(-\bq) \; ,
\end{equation}
where $\eps_{\bk}$ is a single-particle dispersion,
$n_{\bk}(\bq) = 
 \sum_{\sg} c^{\dag}_{\bk-\bq/2,\sg} c_{\bk+\bq/2,\sg}$, and
$V$ is the volume of the system.
Since the Pomeranchuk instability is driven by interactions
with vanishing momentum transfers, that is forward scattering,
we choose a function $f_{\bk\bk'}(\bq)$ which 
contributes only for relatively small momenta $\bq$, and
refer to the above model as the ''\emph{f-model}''.
Clearly this model is adequate only if the Pomeranchuk
instability dominates over other instabilities and
fluctuations in the system. Otherwise it would have to be
supplemented by interactions with large momentum transfers.

For hopping amplitudes $t$, $t'$, and $t''$ between nearest,
next-nearest, and third-nearest neighbors on a square lattice, 
respectively, the dispersion relation is given by
\begin{equation}
 \eps_{\bk} = -2 \left[t (\cos k_{x} + \cos k_{y})
 + 2t'\cos k_{x} \cos k_{y} +t''(\cos 2k_{x} + \cos 2k_{y}) \right]
 \; .
\end{equation}
For a simplified treatment, which however fully captures the 
crucial physics, we consider an interaction of the form
\begin{equation}
 f_{\bk\bk'}(\bq) = u(\bq) + g(\bq) \, d_{\bk} \, d_{\bk'} 
\end{equation}
with $u(\bq) \geq 0$ and $g(\bq) < 0$, and a form factor $d_{\bk}$ 
with $d_{x^2-y^2}$ symmetry, such as $d_{\bk} = \cos k_x - \cos k_y$.
The coupling functions $u(\bq)$ and $g(\bq)$ vanish if 
$|\bq|$ exceeds a certain small momentum cutoff $\Lam$.
This ansatz mimics the effective interaction in the forward 
scattering channel as obtained from renormalization group 
calculations\cite{HM} and perturbation theory\cite{FHR} for the 
two-dimensional Hubbard model near van Hove filling.
The uniform term originates directly from the repulsion between
electrons and suppresses the electronic compressibility of the 
system. The d-wave term drives the Pomeranchuk instability.

The mean-field solution of the f-model has been analyzed
for various choices of parameters in a series of recent 
articles.\cite{KKC,KCOK,YOM}
In the plane spanned by the chemical potential $\mu$ and 
temperature $T$ the symmetry-broken phase is formed below
a dome-shaped transition line $T_c(\mu)$ with a maximal 
transition temperature near van Hove filling. 
The phase transition is usually first order near the edges
of the transition line, that is where $T_c$ is relatively
low, and always second order near its center.\cite{KCOK}
The two tricritical points at the ends of the second order
transition line can be shifted to lower temperatures by a 
sizable uniform repulsion $u$ included in $f_{\bk\bk'}$.\cite{YOM}
For a favorable choice of hopping and interaction parameters,
with a finite $t''$ and $u > 0$, one of the first order edges
is suppressed completely such that a quantum critical point is
realized.
Although quantum critical points are usually prevented by 
first order transitions at low temperatures in the f-model,
the Fermi surface is nevertheless already very soft near the 
transition, such that fluctuations can be expected to be
important.\cite{YOM}

\section{Effective interaction}

In this section we derive and analyze the dynamical effective
interaction for the f-model, which is closely related to the
dynamical \emph{d-wave density correlation} function\cite{Yam}
\begin{equation}
 N_d(\bq,\nu) = -i \int_0^{\infty} dt \; e^{i\nu t} \,
 \bra [ n_d(\bq,t), n_d(-\bq,0) ] \ket \; ,
\end{equation}
where the d-wave density fluctuation operator is defined as
\begin{equation}
 n_d(\bq) = \sum_{\bk} d_{\bk} \, n_{\bk}(\bq) \; .
\end{equation}
and $n_d(\bq,t)$ is the corresponding dynamical operator in 
the Heisenberg picture.
We first analyze $N_d(\bq,\nu)$ and the effective interaction 
within RPA, and then discuss higher order corrections.

The RPA result for the d-wave density correlation function in
the f-model is simply
\begin{equation} \label{ndrpa}
 N_d(\bq,\nu) = 
 \frac{\Pi_d^0(\bq,\nu)}{1 - g(\bq) \, \Pi_d^0(\bq,\nu)}
\end{equation}
with the bare d-wave polarization function
\begin{equation} \label{pid0}
 \Pi_d^0(\bq,\nu) = - \int \frac{d^2p}{(2\pi)^2} \,
 \frac{f(\eps_{\bp+\bq/2}-\mu) - f(\eps_{\bp-\bq/2}-\mu)}
 {\nu + i0^+ - (\eps_{\bp+\bq/2} - \eps_{\bp-\bq/2})} \; 
 d_{\bp}^2 \; .
\end{equation}
The infinitesimal imaginary part in the denominator specifies
that we consider retarded functions.
Note that the coupling $u(\bq)$ does not enter here because
mixed polarization functions with a constant and a d-wave
vertex vanish for small $\bq$.

The RPA effective interaction is defined by the series of chain
diagrams sketched in Fig.\ 1, yielding
\begin{equation} \label{gammarpa}
 \Gam_{\bk\bk'}(\bq,\nu) =
 \frac{u(\bq)}{1 - u(\bq) \, \Pi^0(\bq,\nu)} \, + \,
 \frac{g(\bq)}{1 - g(\bq) \, \Pi_d^0(\bq,\nu)} \, 
 d_{\bk} \, d_{\bk'} \; ,
\end{equation}
where $\Pi^0(\bq,\nu)$ is the conventional polarization function,
defined as $\Pi_d^0(\bq,\nu)$ but without the form factor 
$d_{\bk}$.

Near the Pomeranchuk instability the denominator
$1 - g(\bq) \, \Pi_d^0(\bq,\nu)$ in Eqs.\ (6) and (8)
becomes very small for $\bq \to \b0$ and $\nu \to 0$, 
if $\nu$ vanishes faster than $\bq$,
while $1 - u(\bq) \, \Pi^0(\bq,\nu)$ remains of order one or even
larger. 
The effective interaction is then dominated by the second term
in Eq.\ (8) and can be written as
\begin{equation}
 \Gam_{\bk\bk'}(\bq,\nu) = 
 g \, S_d(\bq,\nu) \, d_{\bk} \, d_{\bk'}
\end{equation}
with $g = g(\b0)$ and the \emph{''dynamical Stoner factor''}
\begin{equation}
 S_d(\bq,\nu) = \frac{1}{1 - g(\bq) \, \Pi_d^0(\bq,\nu)} \; .
\end{equation}
The d-wave density correlation function $N_d(\bq,\nu)$
is also proportional to $S_d(\bq,\nu)$.
In case of a second order transition the static Stoner factor
\begin{equation}
 S_d = \lim_{\bq \to 0} \lim_{\nu \to 0} S_d(\bq,\nu)
\end{equation}
diverges on the transition line. Near the first order transition
obtained typically for low temperatures in the mean-field 
solution of the f-model, $S_d$ is still drastically enhanced,
that is of order ten and more.\cite{YOM}

Within RPA, the dynamical d-wave density fluctuations near the
Pomeranchuk instability and the corresponding singularity of the 
effective interaction are obviously determined by the asymptotic 
behavior of the d-wave polarization function $\Pi_d^0(\bq,\nu)$ 
for small $\bq$ and $\nu$, which we describe in the following; 
for derivations, see Appendix A. 
Analogous formulae for the expansion of the conventional 
polarization function $\Pi^0(\bq,\nu)$ in three dimensions have been
derived already long ago in the context of almost ferromagnetic
metals by Moriya.\cite{Mor}

At zero frequency $\Pi_d^0$ is a real function of $\bq$, which
can be expanded as
\begin{equation} \label{pid0sexp}
 \Pi_d^0(\bq,0) = a(T) + c(T) \, |\bq|^2 + \cO(|\bq|^4)
\end{equation}
for small $\bq$.
The coefficient $a(T)$ is always negative and can be written as
\begin{equation}
 a(T) =
 \int d\eps \, f'(\eps-\mu) \, N_{d^2}(\eps)
\end{equation}
with a weighted density of states $N_{d^2}(\eps) = 
\int \frac{d^2p}{(2\pi)^2} \, \delta(\eps-\eps_{\bp}) \, d_{\bp}^2$.
A low temperature (Sommerfeld) expansion yields
\begin{equation} 
 a(T) = - N_{d^2}(\mu) - \frac{\pi^2}{6} \, 
 N''_{d^2}(\mu) \, T^2 + \cO(T^4) \; ,
\end{equation}
provided that $N_{d^2}(\mu)$ and its second derivative are finite.
The coefficient $c(T)$ can be expressed as
\begin{equation}
 c(T) = \int d\eps \, f'(\eps-\mu) \, 
 \left[ \frac{1}{48} \, N''_{d^2v^2}(\eps) - 
 \frac{1}{16} \, N'_{d^2\Delta\eps}(\eps) \right] \; ,
\end{equation}
where
$N_{d^2\Delta\eps}(\eps) =
 \int \frac{d^2p}{(2\pi)^2} \, \delta(\eps-\eps_{\bp}) \, 
 d_{\bp}^2 \; \Delta\eps_{\bp}$
with the Laplacian
$\Delta = \partial_{p_x}^2 + \partial_{p_y}^2$, and
$N_{d^2v^2}(\eps) = 
 \int \frac{d^2p}{(2\pi)^2} \, \delta(\eps-\eps_{\bp}) \,
 d_{\bp}^2 \, v_{\bp}^2$ 
with $v_{\bp} = |\nabla\eps_{\bp}|$.
Primes denote derivatives with respect to $\eps$.
The sign of $c(T)$ depends on the dispersion and other model
parameters.
In the low temperature limit it will be sufficient to use
\begin{equation}
 c = c(0) = \frac{1}{16} \, N'_{d^2\Delta\eps}(\mu) - 
 \frac{1}{48} \, N''_{d^2v^2}(\mu)
\end{equation}  
For finite frequencies $\Pi_d^0(\bq,\nu)$ is generally complex.
For small $\bq$ and $\nu$ with $\nu/|\bq| \to 0$ its imaginary
part behaves as
\begin{equation}
 \Im\Pi_d^0(\bq,\nu) \to  
 - \, \rho(\hat\bq,T) \, \frac{\nu}{|\bq|} \; ,
\end{equation}
where $\rho(\hat\bq,T) > 0$ depends only on the direction 
$\hat\bq$ of $\bq$.
At low temperatures it is sufficient to use 
$\rho(\hat\bq) = \rho(\hat\bq,0)$, which can be expressed as
\begin{equation} \label{rho0}
 \rho(\hat\bq) = \frac{1}{4\pi} \sum_{\bk_F^0} d_{\bk_F^0}^2 \,
 \frac{1}{v_{\bk_F^0}} \, 
 \frac{1}{|\bt_{\bk_F^0} \cdot {\bf\nabla}_{\bk_F^0} 
 (\hat\bq \cdot \bv_{\bk_F^0})|} \; .
\end{equation}
Here $\bk_F^0$ are points on the Fermi surface satisfying the
condition $\hat\bq \cdot \bv_{\bk_F^0} = 0$ and $\bt_{\bk_F^0}$ 
is a tangential unit vector in $\bk_F^0 \,$. 
Since the velocity $\bv_{\bk_F^0}$ is perpendicular to the Fermi 
surface in $\bk_F^0$ and also perpendicular to $\hat\bq$, the
Fermi surface has to be parallel to $\hat\bq$ in $\bk_F^0$. For
convex reflection symmetric Fermi surfaces in two dimensions 
there are two such points for each given $\hat\bq$, which are
antipodal to each other.

Motivated by the RPA result, but envisaging already
corrections beyond RPA, we parametrize the dynamical Stoner
factor for small $\bq$ and $\nu$, with $\nu/|\bq|$ also small, 
as follows
\begin{equation} \label{sdpar}
 S_d(\bq,\nu) = \frac{1}
 {(\xi_0/\xi)^2 + \xi^2_0 |\bq|^2 -
 i \frac{\nu}{u(\hat\bq) |\bq|}} \; .
\end{equation}
Within RPA, the length scales $\xi_0$ and $\xi$ and the velocity
$u$ are related in a simple manner to the expansion coefficients
of $\Pi_d^0$ and the coupling function $g(\bq)$. Assuming that the
latter can be expanded as $g(\bq) = g + g_2 |\bq|^2 + \dots$
for small $\bq$, we get
\begin{eqnarray}
 \xi_0^2 &=& - g \, c(T) - g_2 \, a(T) \; , \\[2mm]
 (\xi_0/\xi)^2 &=& S_d^{-1} =
 1 - g \, a(T) \; , \\[2mm]
 u(\hat\bq) &=& - \frac{1}{g \, \rho(\hat\bq,T)} \; .
\end{eqnarray}
For $g < 0$ the velocity $u(\hat\bq)$ is always positive.\cite{fn1} 
The static Stoner factor $S_d$ diverges at the Pomeranchuk 
transition, if it is continuous, and the correlation length $\xi$ 
diverges accordingly as $\sqrt{S_d}$. 
The relation for $\xi_0$ makes sense only if the right hand
side is positive. This is guaranteed if we restrict ourselves
to systems where the Pomeranchuk transition is the leading 
instability.
For $-g \, c(T) - g_2 \, a(T) < 0 \,$ a charge density 
wave instability with a wave vector $\bq \neq \b0$ sets in first. 
The parameters $\xi_0$ and $u(\hat\bq)$ remain finite at the
Pomeranchuk transition and do not vary much in its vicinity.
The correlation length $\xi(\delta,T)$ near the transition
depends sensitively on control parameters $\delta$, such as 
the chemical potential, and on the temperature.
If the transition is continuous, $\xi(\delta,T)$ diverges for
$T \to T_c(\delta)$.
Within RPA, $\xi(\delta,T)$ diverges as $(T-T_c)^{-1/2}$ if 
$T_c(\delta) > 0$, and as $T^{-1}$ if $\delta$ is tuned to a 
quantum critical point $\delta_c$.

For small $\bq$ and $\nu$, with $\nu/|\bq|$ also small,
the d-wave density correlation function can be written as
$N_d(\bq,\nu) = - \kappa_d^0 \, S_d(\bq,\nu)$, where
$\kappa_d^0 = - \lim_{\bq \to \b0} \lim_{\nu \to 0} 
\Pi_d^0(\bq,\nu)$ is the static d-wave compressibility.\cite{YOM}
Its imaginary part is then given by
\begin{equation}
 \Im N_d(\bq,\nu) = 
 - \kappa_d^0 \, \frac{\frac{\nu}{u |\bq|}}
 {\big[(\xi_0/\xi)^2 + \xi^2_0 |\bq|^2 \big]^2 +
 \big(\frac{\nu}{u |\bq|} \big)^2}
\end{equation}
For $\xi \gg \xi_0$ and small finite $\bq$ this function 
exhibits a pronounced peak at low frequencies with a steep 
slope for $\nu \to 0$, in agreement with recent numerical 
results for $N_d(\bq,\nu)$ obtained for the t-J model within 
slave boson RPA.\cite{Yam}

We now discuss corrections due to contributions beyond RPA.
The exact density correlation function $N_d(\bq,\nu)$ can be written
in the form of Eq.\ (\ref{ndrpa}), with the full polarization
function $\Pi_d(\bq,\nu)$, which is dressed by interactions, 
instead of the bare one.
Analogously the full effective interaction is given by 
Eq.\ (\ref{gammarpa}) with dressed polarization functions 
$\Pi(\bq,\nu)$ and $\Pi_d(\bq,\nu)$.
Close to a continuous phase transition two types of interaction 
corrections can be distinguished, namely regular interactions
which remain finite at the transition and singular effective
interactions associated with large order parameter fluctuations 
which diverge.

Corrections to RPA and corresponding subleading corrections to
Fermi liquid behavior due to regular interactions, in a generic
stable Fermi liquid regime, have been analyzed thoroughly
in the last few years.\cite{CM}
The low energy behavior of most quantities receives non-analytic 
corrections to Fermi liquid behavior in dimensions $d \leq 3$. 
For example, in two dimensions the spin susceptibility varies 
as $|\bq|$ instead of $|\bq|^2$ for small $\bq$ at $T=0$.
Remarkably, the charge susceptibility for small $\bq$ remains
unaffected. In this case non-analytic contributions appearing 
on the level of single Feynman diagrams cancel systematically 
when all relevant diagrams at a certain order are summed.\cite{CM}
The arguments establishing the cancellation of non-analytic 
corrections for the charge susceptibility can be readily extended 
to our case of a d-wave density instead of the conventional 
density operator.\cite{pcChu}
The crucial point is that a perturbing field coupling to the
d-wave density operator does not alter the singularities in
the polarization function at $\bq = \b0$ and $\bq = 2\bk_F$.
Hence, corrections due to regular interactions may shift the 
parameters with respect to the RPA result for $N_d(\bq,\nu)$
and $\Gamma_{\bk\bk'}(\bq,\nu)$, but they do not yield any
qualitative changes. Within the f-model, one such correction
appears already on mean-field level: the $u$-term in 
$f_{\bk\bk'}(\bq)$ generates a constant (momentum-independent)
Hartree self-energy correction, which renormalizes the 
relation between $\mu$ and density.\cite{YOM}

Near a continuous phase transition order parameter correlations
are strongly renormalized with respect to the mean-field or RPA
result, if the dimensionality of the system is below the upper
critical dimension.\cite{Zin}
These renormalizations are most naturally treated by a 
renormalization group analysis of an effective field theory,
where the order parameter fluctuations are represented by a 
bosonic field. 
The propagator of that field corresponds to the order parameter
correlation function, that is to $N_d(\bq,\nu)$ in our case.
The singular effective interaction between electrons, which is
generated by large order parameter fluctuations, is then
mediated by the bosonic field.
In the bosonic representation, corrections to the RPA result
for the order parameter correlations are due to interactions
of the Bose field.
Close to a continuous Pomeranchuk transition at finite $T_c$ 
these terms are relevant and lead to the classical non-Gaussian 
asymptotic behavior of the Ising universality class in two 
dimensions.
In the following we focus however on the behavior in the 
\emph{quantum} critical regime near the zero temperature critical 
point at $\delta = \delta_c$.
In that regime the upper critical dimension separating Gaussian
from non-Gaussian behavior is $d_c = 4 - z$, where $z$ is the 
dynamical exponent.\cite{Her}
In our case of a charge instability at $\bq=\b0$ one has $z=3$,
in complete analogy to the ferromagnetic quantum critical 
point in itinerant electron systems.\cite{Her}
Hence we have $d_c = 1$, while the dimensionality of our system
is two, such that Gaussian behavior is stable.
However, as first pointed out by Millis,\cite{Mil} the irrelevant
quartic interaction of the order parameter fluctuations changes
the temperature dependence of the correlation length near the
quantum critical point completely compared to the RPA result.
In particular, in a two-dimensional system with $z=3$, the 
correlation length at $\delta_c$ behaves as\cite{Mil}
\begin{equation}
 \xi(\delta_c,T) \propto \frac{1}{\sqrt{T \, |\log T|}}
\end{equation}
instead of the naive $T^{-1}$-divergence.

In summary, the d-wave density correlations and the singular
part of the effective interaction in the quantum critical 
regime can be parametrized by the dynamical Stoner factor
$S_d(\bq,\nu)$ as in Eq.\ (\ref{sdpar}), where the parameters 
$\xi_0$ and $u(\hat\bq)$ do not vary much, while the 
correlation length $\xi(\delta,T)$ depends sensitively
on control parameters $\delta$ and temperature.
The asymptotic behavior of $\xi(\delta,T)$ in the critical
region is strongly influenced by interactions of order parameter 
fluctuations.

\section{Self energy}

We now analyze the low energy behavior of the self-energy in 
the presence of strong d-wave Fermi surface fluctuations.
We compute the self-energy to first order in the singular 
interaction $\Gam_{\bk\bk'}(\bq,\nu)$, non-selfconsistently as
well as selfconsistently, and then discuss the role of vertex 
corrections.

\subsection{Random phase approximation}

To first order in $\Gam$, the self-energy is given by the Fock 
diagram in Fig.\ 2. 
The Hartree term vanishes because the expectation value of
the d-wave density operator vanishes in the symmetric phase.
The Fock diagram yields
\begin{equation} \label{sgrpa}
 \Sg(\bk,i\om_n) = - T \sum_{\nu_n} \int \frac{d^2q}{(2\pi)^2} \,
 \Gam_{\bk\bk}(\bq,i\nu_n) \, G(\bk+\bq,i\om_n+i\nu_n) \, 
 e^{i0^+(\om_n + \nu_n)} \; ,
\end{equation}
where $G$ is the propagator of the interacting system in a
self-consistent perturbation expansion, which is replaced by 
the bare propagator $G_0$ in the non-selfconsistent version.
Note that we have approximated $\Gam_{\bk\bk'}$ with 
$\bk'=\bk+\bq$ by $\Gam_{\bk\bk} \,$, which makes almost no
difference since only small $\bq$ contribute and the effective
interaction does not vary rapidly as a function of $\bk$ and
$\bk'$. 
Analytic continuation to the real frequency axis yields
\begin{eqnarray}
 \Sg(\bk,\om+i0^+) &=& - \, \frac{1}{\pi} \int d\nu 
 \int \frac{d^2q}{(2\pi)^2} \, \big[ 
 b(\nu) \, \Im\Gam_{\bk\bk}(\bq,\nu+i0^+) \, 
 G(\bk+\bq,\nu+\om+i0^+)  \nonumber \\[2mm]
 && \quad - \, f(\nu) \, \Gam_{\bk\bk}(\bq,\nu-\om-i0^+) \, 
 \Im G(\bk+\bq,\nu+i0^+) \big] \; ,
\end{eqnarray}
where $b(\nu) = [e^{\beta\nu}-1]^{-1}$ and 
$f(\nu) = [e^{\beta\nu}+1]^{-1}$ are the Bose and Fermi 
functions, respectively.
It is convenient to focus on the imaginary part of $\Sg$,
from which the real part can be easily recovered via the
Kramers-Kronig relation. 
Using Eq.\ (9) to express the effective interaction by the
dynamical Stoner factor, $\Im\Sg$ can be written as
\begin{equation} \label{imsgrpa}
 \Im\Sg(\bk,\om) = 
 - \, \frac{g \, d_{\bk}^2}{\pi} \int d\nu 
 \int \frac{d^2q}{(2\pi)^2} \, 
 \big[ b(\nu) + f(\nu+\om) \big] \, 
 \Im S_d(\bq,\nu) \, \Im G(\bk+\bq,\om+\nu) \; .
\end{equation}
Here and in the following $G$, $\Sg$, and $S_d$ are retarded functions, 
that is the real frequency axis is approached from above. 
The imaginary part of $S_d$ can be written as 
[see Eq.\ (\ref{sdpar})]
\begin{equation}
 \Im S_d(\bq,\nu) = \frac{u(\hat\bq) \, |\bq| \, \nu}
 {\nu^2 + [(\xi_0/\xi)^2 + (\xi_0 |\bq|)^2]^2 \, 
 [u(\hat\bq) \, |\bq|]^2} \; .
\end{equation}
In the non-selfconsistent calculation one can use
$\Im G_0(\bp,\om) = - \pi \delta(\om - \xi_{\bp})$,
where $\xi_{\bp} = \eps_{\bp} - \mu$,
to perform the $\nu$-integral analytically, which yields
\begin{equation} \label{sgnsc}
 \Im\Sg(\bk,\om) = g \, d_{\bk}^2
 \int \frac{d^2q}{(2\pi)^2} \, \big[ 
 b(\xi_{\bk+\bq}-\om) + f(\xi_{\bk+\bq}) \big] \,
 \Im S_d(\bq,\xi_{\bk+\bq}-\om) \; .
\end{equation}

Since we are interested in the renormalization of low energy
excitations, with $\bk$ close to the Fermi surface, and since
only small momentum transfers $\bq$ contribute to the 
self-energy, it is convenient to introduce a local coordinate
system in momentum space, centered around the Fermi point
$\bk_F$ which is reached from $\bk$ by a normal projection
(see Fig.~3), such that the vector $\bk \!-\! \bk_F$ is 
perpendicular to the Fermi surface in $\bk_F$. 
We can then parametrize $\bk$ by the variable 
$k_r = \pm |\bk \!-\! \bk_F|$, with a positive sign for $\bk$ 
on the exterior side of the Fermi surface, and minus inside.
The momentum transfer $\bq$ can be parametrized by a radial
variable $q_r$ and a tangential variable $q_t$, as shown in
Fig.\ 3.

The excitation energy $\xi_{\bk+\bq}$ appearing in the above
expressions for $\Im\Sg$ can be expanded as
$\xi_{\bk+\bq} = 
 v_{\bk_F} k_r + v_{\bk_F} q_r + \frac{1}{2m_t} \, q_t^2$
for small $\bq$ and $\bk$ near $\bk_F$. 
The parameter $m_t$ is given by the second derivative of $\xi_{\bk}$ 
in tangential direction,
$m_t^{-1} = \partial_{k_t}^2 \xi_{\bk}|_{\bk=\bk_F}$.
The term of order $q_t^2$ has been included since some asymptotic
results are dominated by contributions with $|q_t| \gg |q_r|$.
It is convenient to use
$q'_r = q_r + \frac{1}{2m_t \, v_{\bk_F}} \, q_t^2$ instead of
$q_r$ as integration variable (in addition to $q_t$), since the
excitation energy
$\xi_{\bk+\bq} = v_{\bk_F} (k_r + q'_r)$
is linear in that variable;
the corresponding Jacobi determinant is one. 

\subsubsection{Ground state}

At $T=0$ the combination of Bose and Fermi functions 
contributing to the RPA self-energy reduces to
\begin{equation}
 b(\nu) + f(\nu + \om) = \left\{ \begin{array}{rl}
 - \, \Theta(-\om < \nu < 0) & \; \mbox{for} \quad \om > 0 \\
      \Theta(0 < \nu < -\om) & \; \mbox{for} \quad \om < 0
 \end{array} \right. \; ,
\end{equation}
where $\Theta(.) = 1$, if the inequalities in the argument are
satisfied, and $\Theta(.) = 0$ otherwise.
In the following we restrict to the case $\om > 0$ in
derivations for definiteness, but state final results also for
$\om < 0$. 

At the quantum critical point ($T = 0$, $\xi = \infty$), 
the non-selfconsistent RPA self-energy Eq.\ (\ref{sgnsc})
can be written as
\begin{equation} \label{sgqc}
 \Im\Sg(\bk,\om) = g \, d_{\bk_F}^2
 \int_{-k_r}^{\frac{\om}{v_{\bk_F}} - k_r} \frac{dq'_r}{2\pi} 
 \int \frac{dq_t}{2\pi} \,
 \frac{u(\hat\bq) \, |\bq| \, [\om - v_{\bk_F} (k_r + q'_r)]}
 {[\om - v_{\bk_F} (k_r + q'_r)]^2 + 
 \xi_0^4 \, [u(\hat\bq)]^2 \, |\bq|^6}
\end{equation}
for $\om > 0$. One may impose a cutoff on the $q_t$-integral,
which however does not affect the asymptotic behavior for
small $\om$.

For $\bk = \bk_F$, that is $k_r = 0$, the asymptotic
$\om$-dependence of $\Im\Sg$ can be extracted by introducing
dimensionless variables $\tilde q_r$ and $\tilde q_t$,
which are defined by
$q'_r = (\om/v_{\bk_F}) \, \tilde q_r$ and
$q_t = (\xi_0^2 \, u_{\bk_F})^{-1/3} \om^{1/3} \tilde q_t$,
respectively, where 
$u_{\bk_F} = u(\bt_{\bk_F})$.
For small $\om$ and $k_r = 0$ the above $\bq$-integral is 
dominated by almost tangential $\bq$-vectors, that is
$|q_t| \gg |q_r|$; 
more precisely $|q'_r|$ scales as $|q_t|^3$, and $|q_r|$ 
consequently as $|q_t|^2$.
Hence, we can replace $|\bq|$ by $|q_t|$ and $u(\hat\bq)$ by 
$u_{\bk_F}$. This yields, for $\om \to 0$,
\begin{equation}
 \Im\Sg(\bk_F,\om) \to \frac{g \, d_{\bk_F}^2}{v_{\bk_F}} 
 \, \frac{u_{\bk_F}^{1/3} \, \om^{2/3}}{\xi_0^{4/3}}
 \int_0^1 \frac{d\tilde q_r}{2\pi} 
 \int_{-\infty}^{\infty} \frac{d \tilde q_t}{2\pi} \,
 \frac{|\tilde q_t| \, (1 - \tilde q_r)}
 {(1 - \tilde q_r)^2 + \tilde q_t^6} \; .
\end{equation}
This asymptotic result does not depend on any cutoff.
The definite integral can be done analytically; the result is 
$(4 \sqrt{3} \pi)^{-1}$. For $\om < 0$ one obtains the same 
with $(-\om)^{2/3}$. Hence we have shown that
\begin{equation} \label{sgqckf}
 \Im\Sg(\bk_F,\om) \to
 \frac{g \, d_{\bk_F}^2}{4\sqrt{3}\pi \, v_{\bk_F}} \,
 \frac{u_{\bk_F}^{1/3}}{\xi_0^{4/3}} \, |\om|^{2/3}
\end{equation}
for small $|\om|$.
Note that $\nu = v_{\bk_F} q'_r - \om = 
(\tilde q_r - 1) \, \om$ vanishes faster than $|\bq|$ for 
$\om \to 0$, which justifies our expansion of $S_d(\bq,\nu)$
for small ratios $\nu/|\bq|$.

Not unexpectedly, $\Im\Sg(\bk_F,\om)$ has the same energy
dependence as for the quantum critical points near phase 
separation\cite{CDG} and ferromagnetism\cite{Chu} in two 
dimensions, and also for fermions coupled to a $U(1)$-gauge 
field.\cite{Lee,BM}
In both cases the fluctuation propagator has the same
singularity structure as our Stoner factor $S_d(\bq,\nu)$ for
$\xi = \infty$.
Different is, however, the d-wave form factor making 
$\Im\Sg(\bk_F,\om)$ strongly anisotropic. It is strongest 
near the van Hove points, while the leading terms vanish on
the diagonal of the Brillouin zone.
Subleading terms and contributions from interactions with large
momentum transfers generate at least conventional Fermi liquid
decay rates (of order $\om^2 \log|\om|$) on the diagonal, but
faster decay may be obtained due to higher order processes
which couple different parts of the Fermi surface.

A strongly anisotropic decay rate for single-particle 
excitations following a power-law with exponent $2/3$ has also
been found for an isotropic continuum (not lattice) version
of our model.\cite{OKF} However, that result was obtained
for the symmetry-broken ''nematic'' phase, and the large 
anisotropic decay rate is due to the anisotropy of the nematic
state and its Goldstone modes. 
At the quantum critical point the decay rate of the isotropic
system also obeys a power law with exponent $2/3$, but now
with a constant prefactor over the whole Fermi surface.

For $\bk \neq \bk_F$, that is finite $k_r$, the asymptotic 
behavior of $\Im\Sg(\bk,\om)$, Eq.\ (\ref{sgqc}), is obtained by
rewriting the integral with a dimensionless variable $\tilde q_r$ 
defined by $q'_r + k_r = (\om/v_{\bk_F}) \, \tilde q_r$, 
and $\tilde q_t$ defined by 
$q_t = (\xi_0^2 u)^{-1/3} \om^{1/3} \tilde q_t$.
Here we approximate $u(\hat\bq)$ by a constant $u$ for simplicity.
In the limit $\om \to 0$ one can replace $|\bq|$ by 
$\sqrt{q^2_t + k_r^2}$. 
Carrying out the $\tilde q_r$-integral one then obtains
(including the case $\om < 0$)
\begin{equation}
 \Im\Sg(\bk,\om) = \frac{g \, d_{\bk_F}^2}{2\pi \, v_{\bk_F}} 
 \, \frac{u^{1/3} \, |\om|^{2/3}}{\xi_0^{4/3}}
 \int_0^{\infty} \frac{d \tilde q_t}{2\pi} \,
 \sqrt{\tilde q_t^2 + \kappa^2} \,
 \log\left[1 + 
 \frac{1}{(\tilde q_t^2 + \kappa^2)^3} \right] \; ,
\end{equation}
with $\kappa = (u \xi_0^2/|\om|)^{1/3} \,  k_r$.
Two different types of asymptotic behavior are separated by
the frequency scale
\begin{equation}
 \om_{k_r} = u \xi_0^2 \, |k_r|^3 \; .
\end{equation}
For $|\om| \gg \om_{k_r}$, corresponding to $\kappa \ll 1$, 
one recovers the result Eq.\ (\ref{sgqckf}), 
while for $|\om| \ll \om_{k_r}$ one can expand the integrand 
in $\kappa^{-1}$ and obtains
\begin{equation}
 \Im\Sg(\bk,\om) = 
 \frac{g \, d_{\bk_F}^2}{6\pi^2 \, v_{\bk_F}} \,
 \frac{1}{u \, \xi_0^4 \, k_r^4} \, \om^2 \; .
\end{equation}
In the latter limit momentum transfers normal to the Fermi
surface dominate, and the above result can thus be easily
generalized to a direction dependent $u(\hat\bq)$ replacing
$u$ by $u(\bn_{\bk_F})$.
Note that for small $k_r$, the low frequency behavior of
$\Im\Sg(\bk,\om)$ deviates from $|\om|^{2/3}$-behavior only
below a very small scale of order $|k_r|^3$.
The same crossover behavior has already been obtained 
previously for fermions coupled to a gauge field.\cite{BM}

We now analyze the behavior of $\Im\Sg(\bk_F,\om)$ at $T=0$
in the symmetric phase at some small distance from the quantum 
critical point, where the correlation length $\xi$ is large,
but not infinite. Of interest is in particular at which scale
the $|\om|^{2/3}$-behavior of $\Im\Sg(\bk_F,\om)$ is affected 
by a large finite $\xi$.
Introducing the same dimensionless variables $\tilde q_r$ and
$\tilde q_t$ as for $\xi = \infty$,
one finds that the $\bq$-integral is still
dominated by tangential momentum transfers for small $\om$,
such that we can approximate $|\bq|$ by $|q_t|$ and 
$u(\hat\bq)$ by $u_{\bk_F}$. The $\tilde q_r$-integral can
then be carried out analytically, which yields
\begin{equation}
 \Im\Sg(\bk_F,\om) = 
 \frac{g \, d_{\bk_F}^2}{2\pi \, v_{\bk_F}} 
 \, \frac{u_{\bk_F}^{1/3} \, |\om|^{2/3}}{\xi_0^{4/3}}
 \int_0^{\infty} \frac{d \tilde q_t}{2\pi} \, \tilde q_t \,
 \log\left[ 1 + 
 \frac{1}{\tilde q_t^2 (\tilde q_t^2 + \zeta^2)^2} 
 \right] \; ,
\end{equation}
where $\zeta = (u_{\bk_F} \xi_0^2/|\om|)^{1/3} \xi^{-1}$.
The $\tilde q_t$-integral can be done analytically, leading
however to a rather lengthy expression.
There are two asymptotic regimes, separated by the
characteristic frequency scale
\begin{equation} 
 \om_{\xi} = u_{\bk_F} \xi_0^2/\xi^3 = 
 u_{\bk_F} \xi_0^{-1} S_d^{-3/2} \; .
\end{equation}
For $|\om| \gg \om_{\xi}$ one has $\zeta \ll 1$,
which leads back to the result Eq.\ (\ref{sgqckf}).
For $|\om| \ll \om_{\xi}$, an expansion of the integral 
yields the asymptotic behavior
\begin{equation}
 \Im\Sg(\bk_F,\om) = 
 \frac{g \, d_{\bk_F}^2}{6\pi^2 \, v_{\bk_F}} \,
 \frac{\xi^4}{u_{\bk_F} \xi_0^4} \, \om^2 \, 
 \log\frac{\om_{\xi}}{|\om|} \; .
\end{equation}
Hence, below the scale $\om_{\xi}$ one recovers Fermi liquid
behavior.
Close to the quantum critical point, that is for large $\xi$
and $S_d$, the crossover from $|\om|^{2/3}$ scaling to Fermi
liquid behavior sets in only for very small frequencies, and
the coefficient in front of the asymptotic 
$\om^2 \log\frac{\om_{\xi}}{|\om|}$ law is anomalously large.
A similar non-Fermi to Fermi liquid crossover occurs near 
antiferromagnetic\cite{ACS} and ferromagnetic\cite{Chu}
quantum critical points.
In Fig.\ 4 we show the non-selfconsistent RPA result for
$\Im\Sg(\bk_F,\om)$ in the ground state for various choices of 
$\xi$, as obtained from a numerical integration of the finite
$\xi$ analog of Eq.\ (\ref{sgqc}). The choice of parameters
$\xi_0 = v_{\bk_F} = 1$ corresponds to fixing a length and 
energy scale.
The somewhat arbitrary choice of $u_{\bk_F}$ is not very
critical due to the weak dependence of the results on 
$u_{\bk_F}$.

\subsubsection{Low finite temperature}

At $T > 0$ the self-energy in the non-selfconsistent RPA
approximation, Eq.\ (\ref{sgnsc}), can be written as
\begin{eqnarray} \label{sgnsc'}
 \Im\Sg(\bk,\om) &=& g \, d_{\bk_F}^2
 \int \frac{dq'_r}{2\pi} \int \frac{dq_t}{2\pi} \,
 \big\{ b[v_{\bk_F} (k_r+q'_r) - \om] + 
       f[v_{\bk_F} (k_r+q'_r)] \big\} 
 \nonumber \\[2mm]
 && \times \, 
 \frac{u(\hat\bq) \, |\bq| \, [v_{\bk_F} (k_r+q'_r) - \om]}
 {[v_{\bk_F} (k_r+q'_r) - \om]^2 + 
 [(\xi_0/\xi)^2 + (\xi_0 \, |\bq|)^2]^2 \, [u(\hat\bq) \, |\bq|]^2}
 \; .
\end{eqnarray}
To tackle the asymptotic behavior of this integral for low $T$,
small $\om$ and small $k_r$, it is instructive to consider first 
the special case $k_r = \om = 0$, that is
\begin{eqnarray}
 \Im\Sg(\bk_F,0) &=& g \, d_{\bk_F}^2
 \int \frac{dq'_r}{2\pi} \int \frac{dq_t}{2\pi} \,
 \big[ b(v_{\bk_F} q'_r) + f(v_{\bk_F} q'_r) \big] 
 \nonumber \\[2mm]
 && \times \, 
 \frac{u(\hat\bq) \, |\bq| \, v_{\bk_F} q'_r}
 {(v_{\bk_F} q'_r)^2 + 
 [(\xi_0/\xi)^2 + (\xi_0 \, |\bq|)^2]^2 \, [u(\hat\bq) \, |\bq|]^2}
 \; .
\end{eqnarray}
We introduce dimensionless variables $\tilde q_r$ and $\tilde q_t$
via the relations $q'_r = \xi_0^2 \, \tilde q_r/\xi^3$ 
and $q_t = \tilde q_t/\xi$, respectively.
We now assume that $\xi$ diverges faster than $T^{-1/3}$ for
$T \to 0$, as is indeed the case when we approach the quantum
critical point from the quantum critical region.\cite{Mil}
Then the above integral is dominated by momenta with a small 
ratio $v_{\bk_F} q'_r/T$, such that the Bose function can be 
expanded as $b(v_{\bk_F} q'_r) \to T/(v_{\bk_F} q'_r)$, and the 
Fermi function can be neglected.
Furthermore we can exploit that the integral is dominated by
momenta $\bq$ which are almost tangential to the Fermi surface 
for large $\xi$, since $|q'_r|$ scales as $|q_t|^3$, and hence 
$|q_r|$ as $|q_t|^2$.
We can thus replace $|\bq|$ by $|q_t|$ and $u(\hat\bq)$ by
$u_{\bk_F}$.
We then obtain the simple result
\begin{equation} \label{imsg0}
 \Im\Sg(\bk_F,0) \to 
 \frac{g \, d_{\bk_F}^2}{\xi_0^2} \, T \, \xi 
 \int_{-\infty}^{\infty} \frac{d\tilde q_r}{2\pi} 
 \int_{-\infty}^{\infty} \frac{d \tilde q_t}{2\pi} \;
 \frac{u_{\bk_F} \, |\tilde q_t|}
 {(v_{\bk_F} \tilde q_r)^2 + 
 (1 + \tilde q_t^2)^2 \, (u_{\bk_F} \tilde q_t)^2} =
 \frac{g \, d_{\bk_F}^2}{4 v_{\bk_F} \xi_0^2} \, T \, \xi 
\end{equation}
for $T \to 0$.
A similar contribution of order $T\xi$ has been found already 
earlier for almost antiferromagnetic\cite{VT,ACS} and almost
ferromagnetic\cite{KKI} metals.
Note that $\Im\Sg(\bk_F,0)$ does not obey the same power law
as a function of $T$ as $\Im\Sg(\bk_F,\om)$ as a function of
$\om$ at $T=0$.
The $T^{2/3}$-law proposed in Ref.\ \onlinecite{MRA}, which 
one might expect by identifying $T$ and $\om$ scaling, does
not describe the leading asymptotic behavior at low $T$.

We now generalize the preceding analysis to finite $\om$, which 
shall however be sufficiently small that we can still use the 
expansion of the Bose function and neglect the Fermi function.
To this end we set $\om = v_{\bk_F} x/\xi$, where $x$ is a
dimensionless scaling variable which is kept fixed in the
low temperature limit.
Once again we introduce dimensionless integration variables 
$\tilde q_r$ and $\tilde q_t$, now defined by
$q'_r - \om/v_{\bk_F} = \xi_0^2 \, \tilde q_r/\xi^3$ and
$q_t = \tilde q_t/\xi$, respectively.
For $\xi \to \infty$ we can then replace $q_r$ in 
$|\bq| = \sqrt{q_t^2 + q_r^2}$ by $\om/v_{\bk_F}$, which yields
$|\bq| = \tilde q/\xi$ with 
$\tilde q = \sqrt{\tilde q_t^2 + x^2}$.
The Bose function can again be expanded and the Fermi function
neglected for $T \to 0$, if $\xi$ diverges faster than 
$T^{-1/3}$.
We then obtain
\begin{equation} \label{peak}
 \Im\Sg(\bk_F,\om) \to 
 \frac{g \, d_{\bk_F}^2}{4 v_{\bk_F} \xi_0^2} \, T \, \xi \,
 l(x)
\end{equation}
with a dimensionless scaling function
\begin{equation}
 l(x) = 
 \int_{-\infty}^{\infty} \frac{d\tilde q_r}{2\pi} 
 \int_{-\infty}^{\infty} \frac{d\tilde q_t}{2\pi} \,
 \frac{4 \, v_{\bk_F} \, u(\hat\bq) \, \tilde q}
 {(v_{\bk_F} \tilde q_r)^2 + 
 (1 + \tilde q^2)^2 \, [u(\hat\bq) \, \tilde q]^2}
\end{equation}
The unit vector $\hat\bq$ can be parametrized by $x$ and 
$\tilde q_t$, it does not depend on $\tilde q_r$. 
Doing the elementary $\tilde q_r$-integral, the velocities 
$v_{\bk_F}$ and $u(\hat\bq)$ in the above expression for $l(x)$
drop out completely. 
Carrying out the remaining $\tilde q_t$-integral, we obtain the
simple universal result
\begin{equation} \label{lx}
 l(x) = \frac{1}{\sqrt{1 + x^2}} \; .
\end{equation}
For $\om=0$ one has $l(0) = 1$
and the above special result for $\Im\Sg(\bk_F,0)$ is recovered.
For large $x$ the scaling function decays as $x^{-1}$.
Hence, the contribution from the Bose function singularity to
$\Im\Sg(\bk_F,\om)$ leads to a peak with a height scaling as
$T \xi(T)$ and a width of order $v_{\bk_F}/\xi(T)$.
The product of height and width is thus proportional to $T$.
Since the contribution proportional to the Bose function to
$\Im\Sg(\bk,\om)$, Eq.\ (\ref{sgnsc'}), depends only via the
linear combination $\om - v_{\bk_F} k_r$ on $\om$ and $k_r$,
the right hand side of Eq.~(\ref{peak}) is applicable also for
$\bk \neq \bk_F$, where it yields the contribution from the 
expanded Bose term for 
$\, T \to 0$, $\, \om \to v_{\bk_F} k_r \,$, 
with fixed $x = (\om/v_{\bk_F} - k_r) \, \xi \,$.

The above asymptotic result is entirely due to ''classical'' 
fluctuations, corresponding to the contribution with $\nu_n = 0$ 
to the Matsubara frequency sum in Eq.~(\ref{sgrpa}).
The analytic continuation of that contribution to real 
frequencies reads
\begin{equation}
 \Sg^c(\bk,\om) = - T \int \frac{d^2q}{(2\pi)^2} \,
 \Gam_{\bk\bk}(\bq,0) \, G(\bk+\bq,\om)
\end{equation}
Note that $\Gam_{\bk\bk}(\bq,0)$ is real and does not depend
on the parameter $u(\hat\bq)$. 
For $G = G_0$ one can easily show that indeed 
\begin{equation} \label{sgc0}
 \Im\Sg^c(\bk,\om) \to 
 \frac{g \, d_{\bk_F}^2}{4 v_{\bk_F} \xi_0^2} \, T \, \xi \,
 l[(\om/v_{\bk_F} - k_r)\xi]
\end{equation}
with $l(x)$ from Eq.~(\ref{lx}).
A similar scaling behavior of the self-energy in almost
antiferromagnetic metals has been derived in Ref.~\onlinecite{VT}.

We now split the total self-energy as $\Sg = \Sg^c + \Sg^q$,
where the ''quantum'' contribution is obtained by summing
Matsubara frequencies $\nu_n \neq 0$ in Eq.~(\ref{sgrpa}).
After analytical continuation to real frequencies, $\Im\Sg^c$ 
was obtained from the Bose function singularity $T/\nu$.
To analyze $\Im\Sg^q$ for real frequencies we thus subtract 
$T/\nu$ from the Bose function, that is we replace $b(\nu)$ 
by the regular function 
$\bar b(\nu) = b(\nu) - T/\nu$ in Eq.~(\ref{sgnsc'}).
The asymptotic behavior of $\Im\Sg^q$ at low energy scales
(low frequency and temperature) can be extracted by using
the same dimensionless variables $\tilde q_t$ and $\tilde q_r$
as already in the case $T=0$, and scaling $\om$ as $T$ by
keeping $\tilde\om = \om/T$ fixed in the limit $T \to 0$,
$\om \to 0$.
Asymptotically one can replace $|\bq|$ by $|q_t|$ and 
$u(\hat\bq)$ by $u_{\bk_F}$ as for $T = 0$. 
Furthermore one can neglect $\xi^{-2}$ in the denominator of 
$\Im S_d$ since $\xi^{-2}$ scales to zero faster than $|\bq|^2$.
The $\tilde q_t$-integral can then be carried out analytically,
and we obtain
\begin{equation} \label{sgq0}
 \Im\Sg^q(\bk_F,\om) \to \frac{g \, d_{\bk_F}^2}{v_{\bk_F}} 
 \, \frac{u_{\bk_F}^{1/3} \, |\om|^{2/3}}{\xi_0^{4/3}}
 \, s(\tilde\om) \; ,
\end{equation}
with the universal dimensionless scaling function
\begin{equation} \label{stom}
 s(\tilde\om) = 
 \frac{\sgn(\tilde\om)}{3 \sqrt{3}} 
 \int_{-\infty}^{\infty} \frac{d\tilde q_r}{2\pi} \,
 \left[ \frac{1}{e^{\tilde\om(\tilde q_r - 1)} - 1} -
  \frac{1}{\tilde\om(\tilde q_r - 1)} +
  \frac{1}{e^{\tilde\om \tilde q_r} + 1} \right] \,
 \frac{\tilde q_r - 1}{|1 - \tilde q_r|^{4/3}} \; .
\end{equation}
Note that the above integral converges since the Bose
function pole has been subtracted.
A plot of $s(\tilde\om)$ is shown in Fig.\ 5.
For $|\tilde\om| \to \infty$ the scaling function tends
to $\frac{1}{4\sqrt{3}\pi}$, and one recovers the zero 
temperature result, Eq.~(\ref{sgqckf}).
The convergence to the zero temperature limit is however 
rather slow.  
For small $|\tilde\om|$, $s(\tilde\om)$ is negative and 
proportional to $|\tilde\om|^{-2/3}$, such that
\begin{equation}
 \Im\Sg^q(\bk_F,0) \to 
 \alf \, \frac{g \, d_{\bk_F}^2}{v_{\bk_F}} 
 \, \frac{u_{\bk_F}^{1/3}}{\xi_0^{4/3}} \, T^{2/3} \; ,
\end{equation}
where $\alf \approx - 0.15$ is a numerical constant.
Note that $\Im\Sg^q(\bk_F,0)$ is positive but smaller than 
the absolute value of the classical contribution for low $T$, 
since the latter is proportional to $T \xi$, such that the 
imaginary part of $\Sg^c + \Sg^q$ remains negative, as it should.
For $\bk \neq \bk_F$, that is for finite $k_r$, the momentum
dependence of $\Im\Sg^q(\bk,\om)$ is negligible for
$|\om| \gg \om_{k_r}$, with $\om_{k_r} = u \xi_0^2 |k_r|^3$,
as in the zero temperature case.

To summarize, at low finite $T$ the RPA self-energy is given
by a classical contribution $\Sg^c$ of order $T\xi$, see
Eq.~(\ref{sgc0}), and a quantum contribution $\Sg^q$ of order
$T^{2/3}$ and $|\om|^{2/3}$, which obeys $\om/T$-scaling.
For $|\om| \gg \om_{k_r}$ the latter is described by
Eq.~(\ref{sgq0}).
A similar structure of the self-energy, with a classical part 
and a quantum part obeying $(\om/T)$-scaling, has been obtained 
already earlier for electrons coupled to strong 
ferromagnetic\cite{KKI} or antiferromagnetic\cite{ACS} 
fluctuations in the quantum critical regime. 
In the latter case the self-energy is singular at special hot 
spots on the Fermi surface, and $\Sg^q$ scales with an exponent
$1/2$ instead of $2/3$. 
In Fig.\ 6 we show results for $\Im\Sg(\bk_F,\om)$ as obtained 
from a numerical integration of Eq.~(\ref{sgnsc'}) at various
temperatures. 
The correlation length has been chosen as 
$\xi(T) \propto (T |\log T|)^{-1/2}$, 
that is the temperature dependence derived by Millis.\cite{Mil} 
Analogous results for $\bk \neq \bk_F$ are plotted in Fig.\ 7.
In agreement with the above asymptotic analysis the graphs
show a dispersing finite $T$ structure sitting on top of an 
almost momentum independent background which is proportional
to $|\om|^{2/3}$ for $T \ll |\om|$.

\subsection{Self-consistency}

The self-energy obtained above modifies the propagator $G$ 
strongly at low energy scales. 
We now check the self-consistency of the results obtained in
the preceding subsection, that is we analyze to what extent 
the self-energy obtained from the self-consistent RPA,
Eq.~(\ref{sgrpa}) with a dressed propagator $G$, differs
from the one computed with $G_0$.

\subsubsection{Ground state}

At $T=0$ and for $\om > 0$, $\Im\Sg(\bk,\om)$ in self-consistent 
RPA, Eq.~(\ref{imsgrpa}), can be written as
\begin{equation} \label{imsgrpa2}
 \Im\Sg(\bk,\om) = 
 - \, \frac{g \, d_{\bk}^2}{\pi} \int_0^\om d\eps \,
 \int \frac{dq'_r}{2\pi} \, \int \frac{dq_t}{2\pi} \,   
 \Im S_d(\bq,\eps-\om) \, \Im G(\bk+\bq,\eps) \; .
\end{equation}
We focus on the quantum critical point, $\xi = \infty$,
such that
\begin{equation}
 \Im S_d(\bq,\nu) = \frac{u(\hat\bq) \, |\bq| \, \nu}
 {\nu^2 + \xi_0^4 [u(\hat\bq)]^2 |\bq|^6} \; .
\end{equation}
Motivated by the perturbative results in Sec.\ IV.A,
we assume that the low-frequency behavior of 
$\Im\Sg(\bk_F,\om)$ is given by $C_2 |\om|^{2/3}$, where
$C_2$ is a negative constant, and $\Im\Sg(\bk,\om)$ with
a small distance $k_r$ from the Fermi surface is given by 
the same behavior for frequencies above the small scale 
$\om_{k_r} = u \xi_0^2 \, k_r^3$.
The analytical properties of $\Sg$ in the complex plane 
then dictate
\begin{equation}
 \Re\Sg(\bk_F,\om) = 
 - \frac{\om}{\pi} \, {\cal P} \! 
 \int_{-\infty}^{\infty} d\om' \,
 \frac{\, C_2 \, |\om'|^{2/3}}{(\om - \om') \, \om'} =
 C_1 \, \sgn(\om) |\om|^{2/3} \; ,
\end{equation}
where $\cal P$ denotes the principal value 
and $C_1 = \sqrt{3} \, C_2$.
Anticipating that for small $\om$ and $\bk = \bk_F$ the integral 
in Eq.~(\ref{imsgrpa2}) is dominated by contributions with 
$\eps \gg u \xi_0^2 \, q_r^3$, we can approximate
\begin{equation}
 \Im G(\bk_F+\bq,\eps) = 
 \frac{C_2 \, |\eps|^{2/3}}
 {[\eps - v_{\bk_F} q'_r - C_1 \, \sgn(\eps) |\eps|^{2/3}]^2
  + (C_2 |\eps|^{2/3})^2}
\end{equation}
under the integral.
Scaling out the $\om$-dependence from the integral in
Eq.~(\ref{imsgrpa2}) one finds that the integration variable 
$\eps$ scales as $\om$, $q_t$ as $\om^{1/3}$, and $q'_r$ as 
$\om^{2/3}$. Hence, for small $\om$ one can replace $|\bq|$
by $|q_t|$ and $u(\hat\bq)$ by $u_{\bk_F}$.
This implies that the $q'_r$-integration acts only on
$\Im G$, yielding simply
\begin{equation} \label{qrint}
 \int \frac{dq'_r}{2\pi} \, \Im G(\bk_F+\bq,\eps) =
 - \frac{1}{2v_{\bk_F}} \; .
\end{equation}
This is independent of $C_1$ and $C_2$, which means
that the self-energy drops out completely! Carrying out the
integral over $\eps$ and $q_t$ one then recovers the result
for $\Im\Sg(\bk_F,\om)$ obtained already within the 
non-selfconsistent RPA. 
Hence, the replacement of $G_0$ by $G$ does not affect the
asymptotic low frequency behavior of $\Im\Sg(\bk_F,\om)$
at all, it does not even modify the prefactor.
The same result has already been obtained earlier in the
formally similar problem of fermions coupled to a 
$U(1)$-gauge field,\cite{AIM} and also for antiferromagnetic 
quantum critical points.\cite{ACS}

The above arguments can be easily extended to the case
$\bk \neq \bk_F$ with a small finite $k_r$, since the
latter affects $\Im\Sg(\bk,\om)$ only for frequencies 
below the small scale $\om_{k_r} \propto k_r^3$.

\subsubsection{Low finite temperature}

We first analyze to what extent the peak-shaped classical
contribution to $\Im\Sg(\bk,\om)$ is modified by self-energy 
feedback into $G$. 
As before, we compute this contribution by expanding the
Bose function, $b(\nu) \approx T/\nu$, which yields
\begin{equation}
  \Im\Sg^c(\bk,\om) = 
 - \, \frac{g \, d_{\bk}^2}{\pi} \, T \int \frac{d\nu}{\nu}
 \int \frac{dq'_r}{2\pi} \int \frac{dq_t}{2\pi} \,   
 \Im S_d(\bq,\nu) \, \Im G(\bk+\bq,\om+\nu) \; ,
\end{equation}
where $G$ is the full propagator.
To scale out $\xi$ from $\Im S_d(\bq,\nu)$, we use 
dimensionless variables $\tilde\nu$ and $\tilde q$ 
defined by $\nu = v_{\bk_F} \xi_0^2 \, \tilde\nu/\xi^3$
and $|\bq| = \tilde q/\xi$, respectively, such that
\begin{equation}
 \Im S_d(\bq,\nu) = \frac{\xi^2}{\xi_0^2} \,
 \frac{v_{\bk_F} u(\hat\bq) \, \tilde q \, \tilde\nu}
 {v_{\bk_F}^2 \tilde\nu^2 + (1+\tilde q^2)^2
 [u(\hat\bq) \tilde q \,]^2} \; .
\end{equation}
The inverse propagator can be written as
$G^{-1}(\bk+\bq,\om+\nu) = \om + \nu - v_{\bk_F}(k_r + q'_r)
 - \Sg(\bk+\bq,\om+\nu)$.

Consider first the case $\bk = \bk_F$ and $\om = 0$. 
Then $\Sg(\bk+\bq,\om+\nu)$ can be replaced by 
$\Sg(\bk_F + \bq,0)$
for $T \to 0$, since $\nu$ scales to zero as $\xi^{-3}$.
Motivated by the perturbative result we assume that
$\Im\Sg(\bk,0)$ is of order $T\xi$ for $\bk = \bk_F$ and
decreases for momenta away from $\bk_F$.
Note that $T\xi \propto (\xi \log\xi)^{-1}$, since
$\xi(T) \propto (T \log T)^{-1/2}$.
Then $\nu$ can be neglected completely in the expression
for the propagator, such that
$G^{-1}(\bk_F+\bq,\nu) = 
 - v_{\bk_F} q'_r - \Sg_{\bk_F}(q'_r,0)$, where
$\Sg_{\bk_F}(q'_r,\om) = \Sg(\bk_F + \bq,\om)$.
Now the integral over $\nu$ acts only on $\Im S_d$, yielding
\begin{equation}
 \int \frac{d\nu}{\nu} \, \Im S_d(\bq,\nu) =
 \pi \, \frac{\xi^2}{\xi_0^2} \, \frac{1}{1 + \tilde q^2} =
 \pi \, S_d(\bq,0) \; .
\end{equation}
Writing $\tilde q^2 = \tilde q_r^2 + \tilde q_t^2$, we can
also carry out the integral over $q_t$ and obtain
\begin{equation}
 \int \frac{dq_t}{2\pi} 
 \int \frac{d\nu}{\nu} \, \Im S_d(\bq,\nu) =
 \frac{\pi}{2} \, \frac{\xi}{\xi_0^2} \,
 \frac{1}{\sqrt{1 + \tilde q_r^2}} \; .
\end{equation}
Using $q'_r \approx q_r = \tilde q_r/\xi$ and
collecting all terms we get
\begin{equation}
 \Im\Sg^c(\bk_F,0) = 
 \frac{g \, d_{\bk_F}^2}{2 \, \xi_0^2} \, T \, \xi
 \int \frac{d\tilde q_r}{2\pi} \, 
 \frac{1}{\sqrt{1 + \tilde q_r^2}} \,
 \Im\frac{1}{v_{\bk_F} \tilde q_r + 
 \xi\Sg_{\bk_F}(\tilde q_r/\xi,0)} \; .
\end{equation}
The perturbative result for the self-energy indicates that 
$\Sg_{\bk_F}(\tilde q_r/\xi,0)$ is of order $T\xi$.
For $\xi(T) \propto (T |\log T|)^{-1/2}$ the self-energy 
correction $\xi\Sg_{\bk_F}(\tilde q_r/\xi,0)$ 
in the above integral is then suppressed as $1/\log\xi$ for
$T \to 0$ ($\xi \to \infty$) and can thus be neglected 
compared to $\tilde q_r$, albeit only with logarithmic
accuracy, such that 
$\Im [v_{\bk_F} \tilde q_r + 
 \xi\Sg_{\bk_F}(\tilde q_r/\xi,0)]^{-1}
 \to \pi v_{\bk_F}^{-1} \, \delta(\tilde q_r)$ and we recover 
the perturbative result Eq.~(\ref{imsg0}) for $\Im\Sg(\bk_F,0)$.
More generally, the perturbative result is not modified
(asymptotically) in a self-consistent treatment as long as
$\xi(T)$ increases slower than $T^{-1/2} |\log T|^{1/2}$ for 
$T \to 0$.\cite{pcKat}

For general $\bk$ (at or away from $\bk_F$) and finite $\om$ we 
can still neglect $\nu$ in $G$, such that 
$G^{-1}(\bk+\bq,\om+\nu) = 
 \om - v_{\bk_F} (k_r + q'_r) - \Sg_{\bk_F}(k_r + q'_r,\om)$.
Then $\Im S_d$ can again be integrated independently over
$\nu$ and $q_t$, which yields
\begin{equation} \label{sgcko}
 \Im\Sg^c(\bk,\om) = 
 \frac{g \, d_{\bk_F}^2}{2 \, \xi_0^2} \, T \, \xi
 \int \frac{d\tilde q_r}{2\pi} \, 
 \frac{1}{\sqrt{1 + \tilde q_r^2}} \,
 \Im\frac{1}{v_{\bk_F} \tilde q_r + \xi[v_{\bk_F} k_r -
 \om + \Sg_{\bk_F}(k_r + \tilde q_r/\xi,\om)]} \; .
\end{equation}
The classical contribution to the self-energy feedback in the 
above integral is again suppressed at least as $1/\log\xi$, such 
that only the quantum contribution $\Sg^q$ remains on the right 
hand side.
The latter is not negligible at $\om \neq 0$, it rather
dominates over the bare frequency dependence of $G$ for 
small finite $\om$.
Let us assume that the RPA result for $\Sg^q$ is not 
modified qualitatively by self-consistency, as will be 
indeed verified below.
For $k_r = 0$ or $k_r$ scaling to zero more rapidly than
$|\om|^{1/3}$ one thus has simply
$\Sg_{\bk_F}(k_r + \tilde q_r/\xi,\om) \to 
 \Sg^q_{\bk_F}(\om)$,
where $\Sg^q_{\bk_F}(\om)$ is of order $|\om|^{2/3}$ for
small $\om,T$ with $T \ll |\om|$ and of order $T^{2/3}$ 
for small $\om,T$ with $|\om| \ll T$.
In the latter case $\Sg^q$ is smaller than $\Sg^c$, such that
the self-energy feedback can be neglected completely.
For $|\om - v_{\bk_F}k_r| \ll |\Sg^q_{\bk_F}(\om)|$, the term
$\om - v_{\bk_F}k_r$ can be neglected in Eq.~(\ref{sgcko}), 
such that $\Im\Sg^c(\bk,\om)$ becomes independent of $k_r$ and 
the frequency dependence is given by a scaling function which
decreases monotonically with a dimensionless scaling variable 
proportional to $\xi |\om|^{2/3}$.
Hence, in the self-consistent calculation 
$\Im\Sg^c(\bk,\xi_{\bk})$ is not $k_r$-independent but
rather decreases with increasing $|k_r|$ on a scale of order 
$\xi^{-3/2}$, that is quite rapidly.
For $|\om - v_{\bk_F}k_r| \gg |\Sg^q_{\bk_F}(\om)|$ the
self-energy feedback is negligible and one recovers the
non-selfconsistent RPA result.
In general one will find a crossover between the limiting
cases. In particular, the decrease of $\Im\Sg^c(\bk_F,\om)$ 
as a function of increasing $|\om|$ occurs on a scale of 
order $\xi^{-3/2}$ for the smallest frequencies and then,
more slowly, on the larger scale $\xi^{-1}$.
For $\bk \neq \bk_F$ the self-consistent result for
$\Im\Sg^c(\bk,\om)$ depends on $k_r$ and $\om$ in a more 
complicated way than just via the difference 
$\om - v_{\bk_F} k_r$. Asymptotically it depends only on
$\om$ if $k_r$ scales to zero faster than $|\om|^{2/3}$.

We now check possible modifications of $\Sg^q$ due to 
self-energy feedback at low finite temperatures.
Subtracting the pole from the Bose function, the quantum
contribution to Eq.~(\ref{imsgrpa}) can be written as
\begin{eqnarray}
 \Im\Sg^q(\bk,\om) &=& 
 - \, \frac{g \, d_{\bk}^2}{\pi} \int d\nu 
 \int \frac{dq'_r}{2\pi} \int \frac{dq_t}{2\pi} \,
 \big[ b(\nu) - \frac{T}{\nu} + f(\nu+\om) \big] 
 \nonumber \\[2mm]
 && \times \Im S_d(\bq,\nu) \, \Im G(\bk+\bq,\om+\nu) \; .
\end{eqnarray}
We consider the case $\bk = \bk_F$. 
Frequencies are scaled with temperature, that is 
$\om = T \tilde\om$ and $\nu = T \tilde\nu$. 
At low $T$ and $\om$ one can, once again, replace $|\bq|$ by 
$|q_t|$ and $u(\hat\bq)$ by $u_{\bk_F}$ in $\Im S_d(\bq,\nu)$, 
and neglect $\xi^{-2}$. Then the $q'_r$-integration acts
only on $\Im G$. The inverse propagator can be written as
$G^{-1}(\bk_F+\bq,\om+\nu) = \om + \nu - v_{\bk_F} q'_r -
 \Sg^c_{\bk_F}(q'_r,\om+\nu) - \Sg^q_{\bk_F}(q'_r,\om+\nu)$.
The question now is whether $\Sg^c_{\bk_F}(q'_r,\om+\nu)$
leads to a significant $q'_r$-dependence, which could spoil
the simple result Eq.~(\ref{qrint}) for the $q'_r$-integral
of $\Im G$.
Since $\om+\nu$ scales to zero as $T$, and thus faster than
$\xi^{-3/2}$, one can replace $\Sg^c_{\bk_F}(q'_r,\om+\nu)$
by $\Sg^c_{\bk_F}(q'_r,0)$.
The latter tends to the constant $\Sg^c_{\bk_F}(0,0)$ for
$|q'_r| \ll \xi^{-1}$.
The largest $|q'_r|$ contributing to $\int dq'_r \, \Im G$ 
are proportional to $T \xi$, which indeed scales to zero
faster than $\xi^{-1}$ for $T \to 0$, albeit only by a
factor of order $\log T$.
Hence, to logarithmic accuracy, we can indeed neglect the
$q'_r$-dependence of the self-energy in $\int dq'_r \, \Im G$,
such that the simple formula Eq.~(\ref{qrint}) is still valid,
and the non-selfconsistent result for $\Im\Sg^q(\bk_F,\om)$ is 
confirmed.
The same result is obtained by the same arguments for 
$\bk \neq \bk_F$.

A comparison of the asymptotic result for $\Sg^c$ with a 
numerical evaluation of the self-consistency equations reveals 
that the feedback of $\Sg^c$ into Eq.\ (\ref{sgcko}) becomes 
negligible 
only on extremely small energy scales.
This is not surprising as we have shown that this feedback 
vanishes only logarithmically. 
By contrast, the effects of self-energy feedback into $\Sg^q$ 
are indeed relatively small also for finite temperatures and
frequencies.
In the following, we therefore use the non-selfconsistent RPA
result for $\Sg^q$, but compute $\Sg^c$ from a self-consistent
solution of the complex version of Eq.\ (\ref{sgcko}), that is
\begin{eqnarray} \label{sgcsc}
 \Sg^c(\bk,\om) &=& 
 \frac{g \, d_{\bk_F}^2}{2 \, \xi_0^2} \, T
 \int \frac{dq_r}{2\pi} \, 
 \frac{1}{\sqrt{\xi^{-2} + q_r^2}} \nonumber \\
 &\times& \frac{1}{v_{\bk_F}(q_r + k_r) -
 \om + \Sg^q_{\bk_F}(k_r + q_r,\om) +
 \Sg^c_{\bk_F}(k_r + q_r,\om)} \; .
\end{eqnarray}
This equation can be easily solved numerically by iteration.
Due to the weak radial momentum dependence of $\Sg^q$ one can
neglect the $q_r$-dependence in its argument.
Instead of using the asymptotic scaling form of $\Sg^q$,
which is valid only at sufficiently low $\om$ and $T$, we
compute $\Sg^q$ by integrating Eq.\ (\ref{sgnsc'}) numerically 
and subtracting the classical contribution.
In Figs.\ 8 and 9 we show results for $\Im\Sg(\bk,\om)$ as
obtained by the above procedure. 
One can see that the self-energy feedback suppresses the peak 
generated by classical fluctuations, the effect being of course
stronger for larger $g d_{\bk_F}^2$.

In summary, self-energy feedback in a self-consistent 
calculation affects only the contribution from classical
fluctuations to $\Sg(\bk,\om)$ at $T > 0$.

\subsection{Vertex corrections}

At zero temperature,
vertex corrections and their feedback on the self-energy
have been analyzed in detail by Altshuler et al.\cite{AIM}
for fermions coupled to a $U(1)$-gauge field, which share
many features with our model. They showed that vertex
corrections may lead to moderate finite renormalizations,
but the qualitative behavior of the self-energy, in
particular the power-law with the exponent $2/3$, remains 
unchanged. This was confirmed by a renormalization group analysis 
of the gauge theory, and also for the formally similar case of a 
quantum critical point near phase separation.\cite{CCDM}
The arguments used in the above works can be directly transferred 
to our system and will not be repeated here.

We have not performed a complete analysis of vertex corrections 
at finite temperatures.
By virtue of $\om/T$-scaling,
one may expect that quantum contributions (from finite Matsubara 
frequencies) to vertex corrections behave similarly at zero and
low finite temperatures, and lead to finite renormalizations only.
By contrast, contributions from classical fluctuations at $T>0$ 
have no counterpart at $T=0$ and may thus behave differently.
We have therefore analyzed the first order vertex correction 
(see Fig.\ 10)
due to classical fluctuations in the quantum critical regime.
In the static limit, $\nu = 0$ and $\bq \to \b0$, the classical
contribution to the first order vertex correction is given by
\begin{equation}
 \gam_{\bk_F}^c(\b0,0) \propto 
 T \int d^2q' \, [G(\bk_F+\bq',0)]^2 \, S_d(\bq',0)
\end{equation}
Integrating $S_d(\bq',0)$ over $q'_t$ and using $q'_r = 
\tilde q_r/\xi$, one obtains
\begin{equation}
 \gam_{\bk_F}^c(\b0,0) \propto 
 T \xi^2 \int \frac{d\tilde q_r}{\sqrt{1 + \tilde q_r^2}} \,
 \frac{1}{[v_{\bk_F} \tilde q_r + 
 \xi \Sg_{\bk_F}(\tilde q_r/\xi,0)]^2}
\end{equation}
For $T \to 0$, the self-energy term 
$\xi\Sg_{\bk_F}(\tilde q_r/\xi,0)$ is of order $T \xi^2$ for
small $\tilde q_r$, and vanishes thus logarithmically. 
The above integral diverges thus logarithmically as 
$(T\xi^2)^{-1}$, which cancels precisely the prefactor.
Hence, the classical vertex correction remains finite.

We have not analyzed any higher order vertex corrections.
However, by virtue of the above results and arguments we do
not expect that vertex corrections modify the RPA results for
the self-energy drastically in the quantum critical regime.

\section{Single particle excitations}

The momentum resolved spectral function for single particle
excitations is given by
\begin{equation} \label{specfct}
 A(\bk,\om) = - \frac{1}{\pi} \, \Im 
 \frac{1}{\om - \xi_{\bk} - \Sg(\bk,\om)} =
 \frac{\pi^{-1} \, |\Im\Sg(\bk,\om)|}
 {[\om - \xi_{\bk} - \Re\Sg(\bk,\om)]^2 + [\Im\Sg(\bk,\om)]^2}
 \; ,
\end{equation}
where $\xi_{\bk} = \eps_{\bk} - \mu$ and $\Sg(\bk,\om)$ is
the self-energy computed in the preceding section.
Close to the Fermi surface the self-energy for our model
obeys the symmetry relations
$\Re\Sg_{\bk_F}(-k_r,-\om) = - \Re\Sg_{\bk_F}(k_r,\om)$ and
$\Im\Sg_{\bk_F}(-k_r,-\om) =   \Im\Sg_{\bk_F}(k_r,\om)$,
which implies
$A_{\bk_F}(-k_r,-\om) = A_{\bk_F}(k_r,\om)$,
since $\xi_{\bk} = v_{\bk_F} k_r$ for small $k_r$.

\subsection{Ground state}

In Fig.\ 11 we show results for the spectral function
$A(\bk,\om)$ as a function of $\om$ for several choices
of $k_r$. The underlying self-energy has been computed
at the quantum critical point ($T=0$ and $\xi = \infty$) 
by integrating Eq.~(\ref{sgqc}) numerically.
In the following we discuss the most important features by
using the analytical results from Sec.\ IV.

At the quantum critical point the asymptotic low-energy result 
for the self-energy can be summarized as
\begin{equation}
 \Sg(\bk,\om) \to \Sg_{\bk_F}(\om) = 
 - C_{\bk_F} \, \left[\sgn(\om) + \frac{i}{\sqrt{3}} \right] \, 
 |\om|^{2/3} \;
 \mbox{, where} \; \;
 C_{\bk_F} = \frac{|g| \, d_{\bk_F}^2}{4\pi \, v_{\bk_F}} \,
 \frac{u_{\bk_F}^{1/3}}{\xi_0^{4/3}} \; .
\end{equation}
Strictly speaking this simple $k_r$-independent behavior 
is valid only for $|\om| \gg \om_{k_r}$, but the scale 
$\om_{k_r}$ is proportional to $k_r^3$ and thus tiny for $\bk$ 
near the Fermi surface.
The prefactor $C_{\bk_F}$ depends strongly on the position
of $\bk_F$.
It decreases rapidly for $\bk_F$ near the Brillouin zone
diagonal, where $d_{\bk_F}$ vanishes, while it is
enhanced near the van Hove points, where $v_{\bk_F}$
becomes small. The $\bk_F$-dependence of $u_{\bk_F}^{1/3}$
is comparatively weak.
The competition between $\om$ and the self-energy in the 
denominator of $A(\bk,\om)$, Eq.~(\ref{specfct}), leads to
the characteristic energy scale
\begin{equation} \label{omkf}
 \om_c = C_{\bk_F}^3 \propto 
 \frac{d_{\bk_F}^6}{v_{\bk_F}^3} \; ,
\end{equation}
which obviously varies very strongly with $\bk_F$.

For fixed $\bk$ with $|\xi_{\bk}| \gg \om_c$ the spectral 
function $A(\bk,\om)$ has almost Lorentzian shape as a function 
of $\om$, with a maximum near $\om = \xi_{\bk}$ and a width of
order $C_{\bk_F} |\xi_{\bk}|^{2/3}$.
The life-time broadening thus decreases more slowly than the
excitation energy $\xi_{\bk}$, as $\bk$ approaches the Fermi 
surface, such that no well-defined quasi-particles exist.
For $|\xi_{\bk}| \approx \om_c$ the maximum of $A(\bk,\om)$
is shifted strongly away from $\xi_{\bk}$ and the width is
of order of the peak energy. 
For $|\xi_{\bk}| \ll \om_c$ one can neglect $\om$ compared
to $\Re\Sg(\bk,\om)$ in Eq.~(\ref{specfct}), and $A(\bk,\om)$
is now peaked at $\om = \bar\xi_{\bk}$, with the renormalized
energy scale
\begin{equation}
 \bar\xi_{\bk} = 
 \sgn(\xi_{\bk}) \, (C_{\bk_F}^{-1} |\xi_{\bk}|)^{3/2} 
 \propto k_r^{3/2} \; .
\end{equation}
Extracting a dispersion relation from the momentum dependence
of the peak position in $A(\bk,\om)$ one thus obtains a flat 
band with a vanishing slope near the Fermi surface.
The width of the peak centered around $\bar\xi_{\bk}$ is
of order $C_{\bk_F} |\bar\xi_{\bk}|^{2/3} = |\xi_{\bk}|$ 
and thus linear in $k_r$. 
For $\bk = \bk_F$ the spectral function diverges as
$|\om|^{-2/3}$ for $\om \to 0$.

Momentum scans of $A(\bk,\om)$ perpendicular to the Fermi
surface at fixed $\om$ lead to Lorentzian peaks centered
around 
$k_r = \frac{1}{v_{\bk_F}} 
 [\om + C_{\bk_F} \sgn(\om)|\om|^{2/3}] \,$.
Some such scans are shown in Fig.\ 12 for various choices
of $\om$. 
The integral
\begin{equation}
 \int \frac{dk_r}{2\pi} \, A(\bk,\om) = 
 \frac{1}{2\pi \, v_{\bk_F}}
\end{equation}
does not depend on the self-energy. The $\bk$-integrated
density of states near the Fermi level remains therefore 
unrenormalized, that is, it is determined by the bare
Fermi velocity.

\subsection{Low finite temperature}

At low finite $T$ the results for the self-energy can be
summarized as follows. The total self-energy is a sum of
two distinct contributions, $\Sg = \Sg^c + \Sg^q$, where
$\Sg^c$ is due to classical, and $\Sg^q$ due to quantum
fluctuations. 
The quantum contribution $\Sg^q$ can be computed from
non-selfconsistent RPA and obeys $(\om/T)$-scaling at very
low $\om$ and $T$, see Eq.~(\ref{sgq0}). Its dependence
on $k_r$ is very weak.
The classical contribution $\Sg^c$ is affected more strongly
by self-energy feedback, especially by feedback of $\Sg^q$. We 
compute $\Sg^c$ by solving Eq.\ (\ref{sgcsc}) self-consistently.
The classical part violates $(\om/T)$-scaling and depends
significantly on $k_r$.

The most significant temperature effect is that 
$\Im\Sg(\bk_F,0)$ increases quickly from zero to sizable
finite values upon increasing $T$. For small $T$ near
the quantum critical point we obtained
\begin{equation}
 \Im\Sg(\bk_F,0) \to
 \frac{g \, d_{\bk_F}^2}{4 v_{\bk_F} \xi_0^2} \, T \, \xi
 \propto \frac{d_{\bk_F}^2}{v_{\bk_F}} \,
 \sqrt{\frac{T}{|\log T|}}
\end{equation}
both in the non-selfconsistent and self-consistent
calculation. 
This cuts off the divergence of $A(\bk_F,\om)$ for $\om \to 0$
occurring at zero temperature and replaces it by a maximum
of order $\sqrt{|\log T|/T}$.
Note that $\Im\Sg(\bk_F,0)$ vanishes much slower with $T$
than in conventional Fermi liquids, where one has $T^2$ (in 3D) 
or $T^2 |\log T|$ (in 2D) behavior.

In Fig.~13 we show results for $A(\bk,\om)$ as a function of 
$\om$ for various choices of $k_r$ at a fixed low temperature.
The underlying self-energy has been computed by the procedure
described at the end of Sec.\ IV.B.
The most striking differences compared to the ground state
results (Fig.~11) are of course seen for $\bk$ near $\bk_F$, 
where the peaks are now much broader.
Note also the steep shoulder near $\om = 0$ for the spectra
with $\bk$ near $\bk_F$ at strong coupling (lower panel).  
The complementary view, $A(\bk,\om)$ as a function of $k_r$
for various fixed $\om$, is shown in Fig.~14. 
Here the line shape resembles a Lorentzian function with a
relatively large width.

\section{Conclusion}

In summary, we have presented a detailed analysis of quantum 
critical fluctuations and their effect on single-particle
excitations in a two-dimensional electron system close to
a d-wave Pomeranchuk instability.
The fluctuations can be viewed as long-wavelength density 
fluctuations with a d-wave form factor and also as d-wave
shaped fluctuations of the Fermi surface.
They lead to a strong singularity in the dynamical d-wave 
density correlation function at small momenta and 
frequencies, and to singular forward scattering. 
The momentum and energy dependence of the singularity is
captured essentially correctly by a Gaussian theory (RPA);
interactions of fluctuations modify only the temperature
dependence of the correlation length $\xi(T)$.

Single-particle excitations are strongly affected by the
fluctuations.
We have analyzed the electron self-energy $\Sg(\bk,\om)$ 
within plain and self-consistent RPA, focussing especially 
on the low-energy behavior in the quantum critical regime.
The dominant contributions due to singular forward scattering 
are proportional to $d_{\bk}^2$, where $d_{\bk}$ is a form
factor with d-wave symmetry, such as $\cos k_x - \cos k_y$.
For $\bk$ near the Fermi surface this leads to a strong 
tangential momentum dependence of $\Sg(\bk,\om)$.
The singular contributions vanish on the diagonal of the 
Brillouin zone, and have the largest amplitude near the 
van Hove points.
By constrast, the momentum dependence of $\Sg(\bk,\om)$ 
perpendicular to the Fermi surface is much weaker at low
temperatures.
Hence, momentum scans of the spectral function perpendicular
to the Fermi surface have almost Lorentzian line shape.

At the quantum critical point, the real and imaginary parts
of the self-energy scale as $|\om|^{2/3}$ with energy. 
This leads to a complete destruction of quasi-particles
near the Fermi surface except on the Brillouin zone
diagonal, due to the prefactor $d_{\bk}^2$.
The dispersion of the maxima of the spectral function
$A(\bk,\om)$ flattens strongly for momenta $\bk$ near the 
Fermi surface away from the zone diagonal.
On the other hand, the momentum integrated density of 
states is not renormalized significantly by the Fermi surface 
fluctuations.

In the quantum critical regime at $T>0$ the self-energy
consists of a ''classical'' and a ''quantum'' part with
very different dependences on $T$ and $\om$. 
The classical part, which is due to classical fluctuations,
dominates at $\om = 0$, where it yields a contribution
proportional to $T\xi(T)$ to $\Im\Sg(\bk_F,\om)$. 
The quantum part is generated by quantum fluctuations and 
obeys $(\om/T)$-scaling in the quantum critical regime.

We finally discuss whether soft Fermi surfaces and critical 
Fermi surface fluctuations could play a role in cuprate 
superconductors.
Due to the coupling of electron and lattice degrees of freedom
a symmetry-breaking Fermi surface deformation is generally 
accompanied by a lattice distortion, and vice versa. 
Structural transitions which reduce the lattice symmetry of 
the cuprate-planes are quite frequent in cuprates.
Close to a Pomeranchuk instability of the electronic system, 
electronic properties can be expected to react unusually 
strongly to slight lattice distortions which break the
symmetry of the electronic system explicitly. 
Such ''overreactions'' of electronic properties have indeed been 
observed early on in several cuprate compounds.\cite{Axe,Bue} 
In particular, a slight orthorhombicity of the lattice structure 
would lead to a relatively strong orthorhombic distortion of the 
Fermi surface. Yamase and Kohno \cite{YK2} invoked this idea
already a few years ago to explain peculiarities of magnetic 
excitations in cuprates. 
Recent experiments on YBCO have established a remarkably 
strong in-plane anisotropy of electronic and magnetic 
properties,\cite{Lu,Ando,Hin}
although the structural anisotropy of the $\rm CuO_2$-planes,
which is induced indirectly by the $\rm CuO$-chains between the 
planes in that material, is relatively modest.\cite{And}
Fermi surface softening with d-wave symmetry due to forward
scattering interactions can naturally amplify the effect of a 
weak or moderate structural anisotropy.

Since the Pomeranchuk instability breaks the orientational
symmetry of the lattice, it is natural to compare with the
mechanisms and consequences of stripe formation, which
has been extensively discussed in the context of cuprate
superconductors.\cite{stripes}
Static stripes also break the translation invariance in 
addition to the orientational symmetry, and their formation
requires interactions with large momentum transfers, such 
as antiferromagnetic interactions.
Many experimental observations, which have been attributed
to static or fluctuating stripes,\cite{KBX} actually provide 
evidence only for a tendency to orientational, not translational,
symmetry-breaking, and could therefore be described equally 
well by a (incipient) Pomeranchuk instability.

Strong Fermi surface fluctuations could be at least partially
responsible for the non-Fermi liquid behavior observed in the
''strange metal'' regime of cuprate superconductors near 
optimal doping. In our model calculation we have obtained a
strongly anisotropic anomalously large decay rate for
single-particle excitations and a flattening of the dispersion
relation near the Fermi surface away from the nodal direction.
The properties of single-particle excitations in various
cuprate compounds have been investigated in considerable 
detail by numerous angular resolved photoemission 
experiments.\cite{DSH}
Extended flat bands in the van Hove region have been observed
by various groups already in the early 1990s.\cite{Abr,Des,Gof}
Large anisotropic decay rates have been extracted from the 
linewidth of low-energy peaks in the photoemission spectra
observed in optimally doped cuprates, using in particular 
momentum scans perpendicular to the Fermi surface at various 
fixed energies.\cite{Val,Kam} 
The line shape of these scans is almost Lorentzian, which is 
consistent with our results.
However, the frequency and temperature dependence of the
self-energy extracted from the experimental raw-data differs
from what we obtained from d-wave Fermi surface fluctuations.
On the other hand, the functional form of the self-energy
chosen by the experimentalists may not be the only way to
achieve consistency with the photoemission data. In spite
of the impressive progress in this experimental technique, the
accuracy of such subtle properties as the energy and temperature 
dependence of the electron self-energy is still limited by 
resolution and background problems.

Concerning transport, an anisotropic scattering rate with nodes
on the Brillouin zone diagonal can very naturally account for 
the pronounced anisotropy between the intra- and inter-plane 
mobility of charge carriers, as pointed out by Ioffe and 
Millis\cite{IM} in their phenomenological ''cold spot'' scenario. 
According to their idea, the intra-plane transport is dominated 
by quasi-particles with a long life-time near the diagonal of 
the Brillouin zone, while these carriers are not available for
inter-plane transport, since transverse hopping amplitudes
vanish on the diagonal. 

A spin dependent Pomeranchuk instability was recently invoked 
to explain a new phase observed in ultrapure crystals of the 
layered ruthenate metal $\rm Sr_3 Ru_2 O_7$,\cite{GGX}
and also to account for a puzzling phase transition in
$\rm U Ru_2 Si_2$.\cite{VZ} 
For a broader comparison with experimental data for cuprate
superconductors and other layered materials, which might 
undergo or be close to a symmetry-breaking Fermi surface
deformation, it will be useful to compute experimentally
accessible response functions in the presence of strong
Fermi surface fluctuations.

\acknowledgments

We thank S. Andergassen, A. Chubukov, C. Di Castro, D. Rohe, 
M. Vojta, H. Yamase, and especially C. Castellani and 
A. Katanin for very valuable discussions.

\appendix

\section{Bare polarization function}

In this appendix we derive the asymptotic expressions for
the bare polarization function $\Pi_d^0(\bq,\nu)$ defined in
Eq.\ (\ref{pid0}) for small $\bq$ and $\nu$.
The derivation is valid for arbitrary form factors $d_{\bp}$
with any symmetry (not only d-wave), in particular also for
the special case $d_{\bp} = 1$, for which $\Pi_d^0$ reduces to
the conventional polarization function $\Pi^0$.

At zero frequency, $\Pi_d^0$ can be written as
\begin{equation} \label{pid0s}
 \Pi_d^0(\bq,0) = \int \frac{d^2p}{(2\pi)^2} \,
 \frac{f(\xi_{\bp+\bq/2}) - f(\xi_{\bp-\bq/2})}
 {\eps_{\bp+\bq/2} - \eps_{\bp-\bq/2}} \; 
 d_{\bp}^2 \; .
\end{equation}
with $\xi_{\bp} = \eps_{\bp} - \mu$. The numerator and the
denominator of the fraction under the above integral are odd
functions of $\bq$, and the integrand is thus an even function.
The denominator can be expanded as
\begin{equation}
 \eps_{\bp+\bq/2} - \eps_{\bp-\bq/2} = 
 \bq \cdot \bv_{\bp} + 
 \frac{1}{24} \sum_{j_1,j_2,j_3} \frac{\partial^3\eps_{\bp}}
 {\partial p_{j_1} \partial p_{j_2} \partial p_{j_3}} \,
 q_{j_1} q_{j_2} q_{j_3} + \cO(|\bq|^5)
\end{equation}
with $\bv_{\bp} = \nabla_{\!\bp} \eps_{\bp}$
and $j_1,j_2,j_3$ each running over the two possible directions
$x$ and $y$ (in two dimensions).
The numerator is expanded as
\begin{equation}
 f(\xi_{\bp+\bq/2}) - f(\xi_{\bp-\bq/2}) = 
 \bq \cdot \nabla_{\!\bp} f(\xi_{\bp}) + 
 \frac{1}{24} \sum_{j_1,j_2,j_3} \frac{\partial^3 f(\xi_{\bp})}
 {\partial p_{j_1} \partial p_{j_2} \partial p_{j_3}} \,
 q_{j_1} q_{j_2} q_{j_3} + \cO(|\bq|^5) \; .
\end{equation}
Using $\nabla_{\!\bp} f(\xi_{\bp}) = 
f'(\xi_{\bp}) \, \bv_{\bp}$
and
\begin{eqnarray}
 \sum_{j_1,j_2,j_3} \frac{\partial^3 f(\xi_{\bp})}
 {\partial p_{j_1} \partial p_{j_2} \partial p_{j_3}} \,
 q_{j_1} q_{j_2} q_{j_3} &=&
 f'''(\xi_{\bp}) \, (\bq \cdot \bv_{\bp})^3 
 + 3 f''(\xi_{\bp}) \, (\bq \cdot \bv_{\bp})
 \sum_{j_1,j_2} \frac{\partial^2 \eps_{\bp}}
 {\partial p_{j_1} \partial p_{j_2}} \, q_{j_1} q_{j_2}
 \nonumber \\ && + \,
 f'(\xi_{\bp}) \sum_{j_1,j_2,j_3} \frac{\partial^3\eps_{\bp}}
 {\partial p_{j_1} \partial p_{j_2} \partial p_{j_3}} \,
 q_{j_1} q_{j_2} q_{j_3}
\end{eqnarray}
one obtains
\begin{equation}
 \frac{f(\xi_{\bp+\bq/2}) - f(\xi_{\bp-\bq/2})}
 {\eps_{\bp+\bq/2} - \eps_{\bp-\bq/2}} =
 f'(\xi_{\bp}) + \frac{1}{8} f''(\xi_{\bp}) 
 \sum_{j_1,j_2} \frac{\partial^2 \eps_{\bp}}
 {\partial p_{j_1} \partial p_{j_2}} \, q_{j_1} q_{j_2} +
 \frac{1}{24} f'''(\xi_{\bp}) \, (\bq \cdot \bv_{\bp})^2 +
 \cO(|\bq|^4)
\end{equation}
Note that terms involving third order derivates of $\eps_{\bp}$ 
have cancelled.
Inserting this expansion in Eq.\ (\ref{pid0s}) and using the
point group symmetries of the square lattice one obtains
\begin{equation}
 \Pi_d^0(\bq,0) = \int \frac{d^2p}{(2\pi)^2} \left[
 f'(\xi_{\bp}) + 
 \frac{1}{16} f''(\xi_{\bp}) \, \Delta\eps_{\bp} \, |\bq|^2 +
 \frac{1}{48} f'''(\xi_{\bp}) \, v_{\bp}^2 \, |\bq|^2
 \right] d_{\bp}^2 + \cO(|\bq|^4) \; ,
\end{equation}
which establishes Eq.\ (\ref{pid0sexp}) and the formulae for the 
expansion coefficients $a(T)$ and $c(T)$ presented in Sec.\ III.
In the special case $d_{\bk} = 1$ the term involving $f''$ in
the above equation can be rewritten in the form of the term
proportional to $f'''$ by a partial integration, yielding the
simplified formula for the conventional polarization function
\begin{equation}
 \Pi^0(\bq,0) = \int \frac{d^2p}{(2\pi)^2} \left[
 f'(\xi_{\bp}) -
 \frac{1}{24} f'''(\xi_{\bp}) \, v_{\bp}^2 \, |\bq|^2
 \right] + \cO(|\bq|^4) \; .
\end{equation}

We now derive the frequency and momentum dependence of 
$\Pi_d^0(\bq,\nu)$ for small $\bq$ and $\nu$ to leading order
in $\bq$ and $\nu$. 
In the limit $\bq \to 0$, $\nu \to 0$ the polarization function
depends only via the ratio $s = \nu/|\bq|$ and the unit vector
$\hat\bq = \bq/|\bq|$ on $\bq$ and $\nu$:
\begin{equation}
 \Pi_d^0(\bq,\nu) \, \to \,
 - \int \frac{d^2p}{(2\pi)^2} \, f'(\xi_{\bp}) \,
 \frac{\bv_{\bp} \cdot \hat\bq}
 {s + i0^+ - \bv_{\bp} \cdot \hat\bq} \, d_{\bp}^2 \; .
\end{equation}
In the low temperature limit only momenta on the Fermi surface
contribute to the above integral, since then
$f'(\xi_{\bp}) \to - \delta(\xi_{\bp})$.
The real part is an even function of $s$ which tends to $a(T)$ 
in the limit $s \to 0$, with corrections of order $s^2$.
The imaginary part has the simple form
\begin{equation} \label{impd0}
 \Im\Pi_d^0(\bq,\nu) \, \to \,
 \int \frac{d^2p}{4\pi} \, f'(\xi_{\bp}) \, d_{\bp}^2 \,
 s \, \delta(s - \bv_{\bp} \cdot \hat\bq) \; .
\end{equation}
For $\bq,\nu \to 0$ and small $s$ this simplifies further to
$\Im\Pi_d^0(\bq,\nu) \, \to \, - \rho(\hat\bq,T) \, s$ with
\begin{equation}
 \rho(\hat\bq,T) = 
 - \int \frac{d^2p}{4\pi} \, f'(\xi_{\bp}) \, d_{\bp}^2 \,
 \delta(\bv_{\bp} \cdot \hat\bq) \; .
\end{equation}
At $T=0$ the integrand in Eq.\ (\ref{impd0}) contains two
$\delta$-functions. The two-dimensional momentum integral can 
then be carried out, yielding
\begin{equation}
 \Im\Pi_d^0(\bq,\nu) \, \to \,
 - \frac{s}{4\pi} \sum_{\bk_F^0} d_{\bk_F^0}^2 \,
 \frac{1}{v_{\bk_F^0}} \, 
 \frac{1}{|\bt_{\bk_F^0} \cdot {\bf\nabla}_{\bk_F^0} 
 (\hat\bq \cdot \bv_{\bk_F^0})|} \; ,
\end{equation}
where $\bt_{\bk_F^0}$ is a unit vector tangential to the
Fermi surface in $\bk_F^0$, 
and the sum runs over momenta on the Fermi surface which
satisfy the equation $\bv_{\bk_F^0} \cdot \hat\bq = s$.
The formula (\ref{rho0}) for $\rho(\hat\bq,T)$ at $T=0$
follows directly.


\vfill\eject


\begin{figure}
\center
\epsfig{file=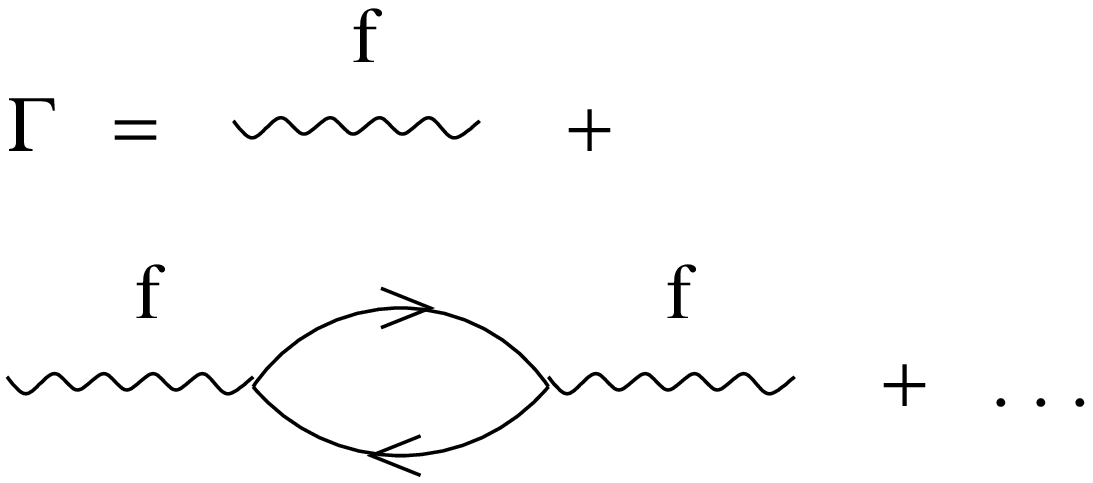,width=8cm}
\caption{Series of bubble chains contributing to the effective
 interaction $\Gam$.} 
\end{figure}

\begin{figure}
\center
\epsfig{file=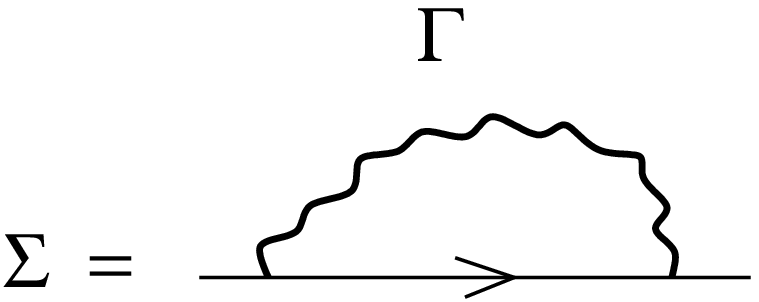,width=5cm}
\caption{Fock diagram relating the self-energy $\Sg$ to the
 effective interaction $\Gam$.} 
\end{figure}

\begin{figure}
\center
\epsfig{file=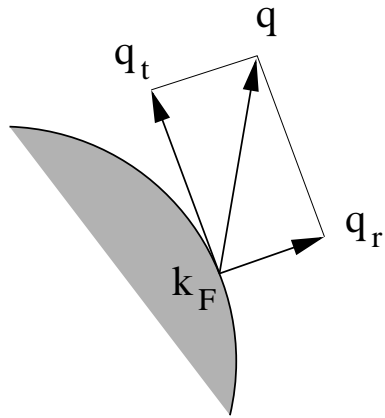,width=5cm}
\caption{Decomposition of momentum transfers $\bq$ in normal 
 and tangential components relative to the Fermi surface in 
 $\bk_F$.} 
\end{figure}

\begin{figure}
\center
\epsfig{file=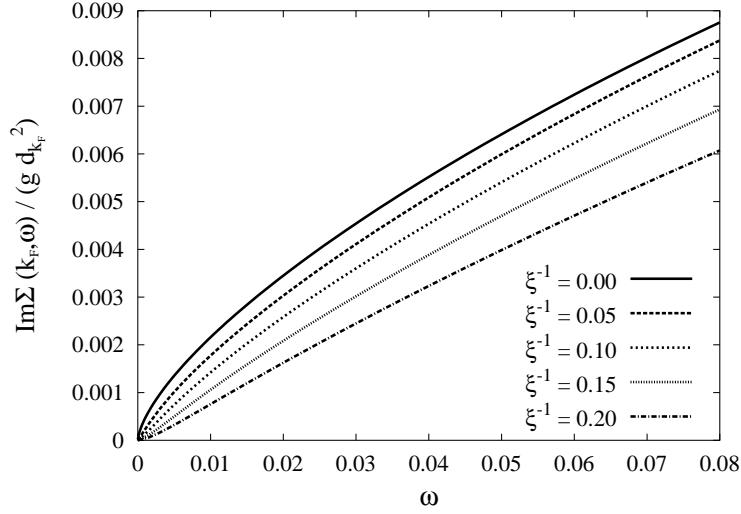,width=10cm}
\caption{Non-selfconsistent RPA result for the
 imaginary part of the self-energy $\Im\Sg(\bk_F,\om)$, divided
 by $g d_{\bk_F}^2$, as a function of $\om$ in the ground 
 state ($T=0$) for various choices of the correlation length 
 $\xi$; the other relevant parameters are 
 $\xi_0 = v_{\bk_F} = u_{\bk_F} = 1$.} 
\end{figure}

\begin{figure}
\center
\epsfig{file=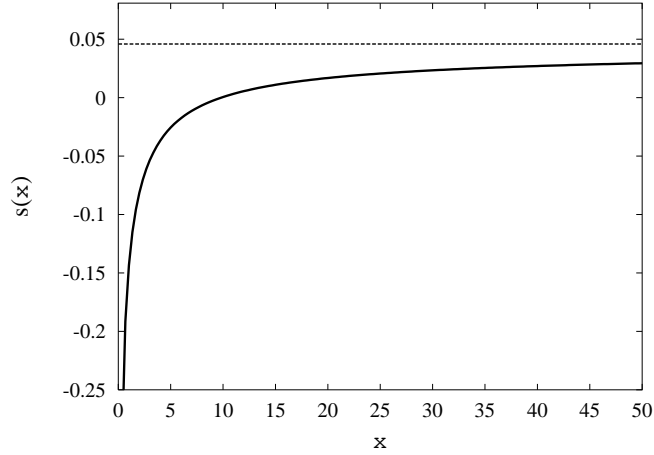,width=9cm}
\caption{Scaling function $s(x)$ describing the 
 asymptotic behavior of the quantum contribution to 
 $\Im\Sg(\bk_F,\om)$ at low $\om$ and $T$; the horizontal line
 indicates the asymptotic value of $s(x)$ for large $x$.}
\end{figure}

\begin{figure}
\center
\epsfig{file=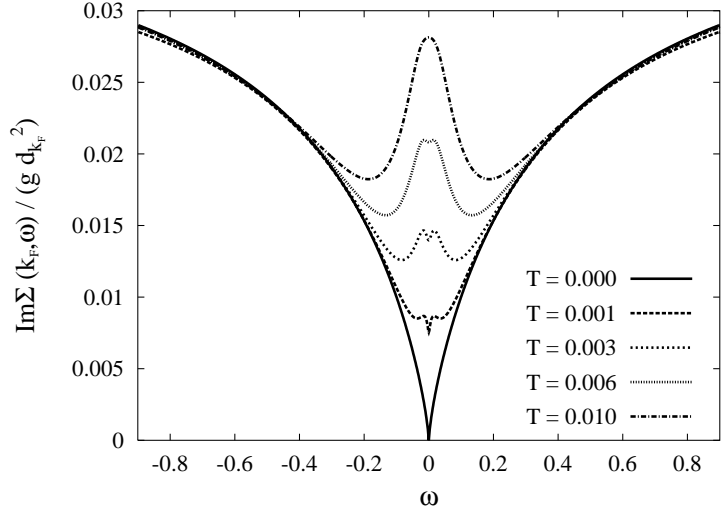,width=10cm}
\caption{Non-selfconsistent RPA result for the
 imaginary part of the self-energy $\Im\Sg(\bk_F,\om)$, divided
 by $g d_{\bk_F}^2$, as a function of $\om$ for several
 temperatures $T$ with a correlation length 
 $\xi(T) = 3/|T \log T|^{1/2}$;
 the other relevant parameters are $\xi_0 = v_{\bk_F} = u = 1$.} 
\end{figure}

\begin{figure}
\center
\epsfig{file=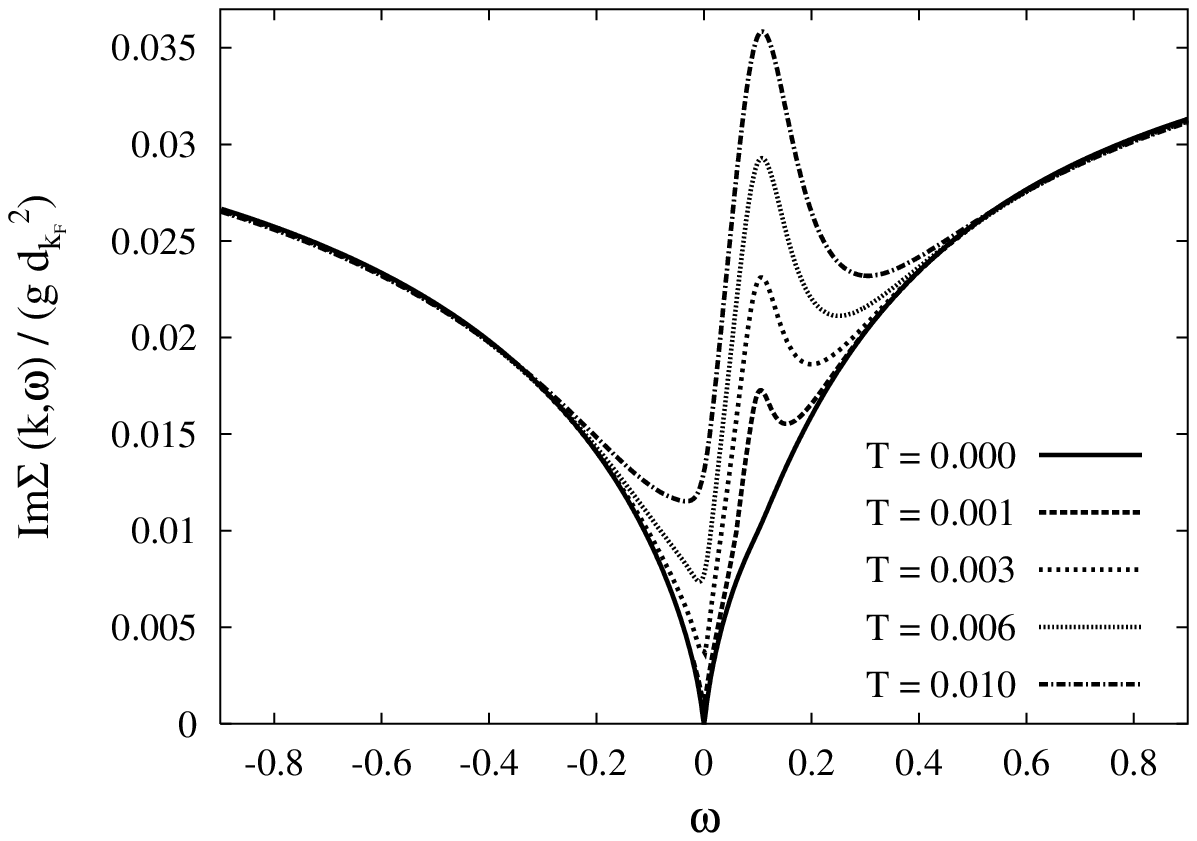,width=10cm}
\caption{Non-selfconsistent RPA result for $\Im\Sg(\bk,\om)$ as
 in Fig.\ 6, but now for $k_r = 0.1$, that is $\bk \neq \bk_F$.} 
\end{figure}

\begin{figure}
\center
\epsfig{file=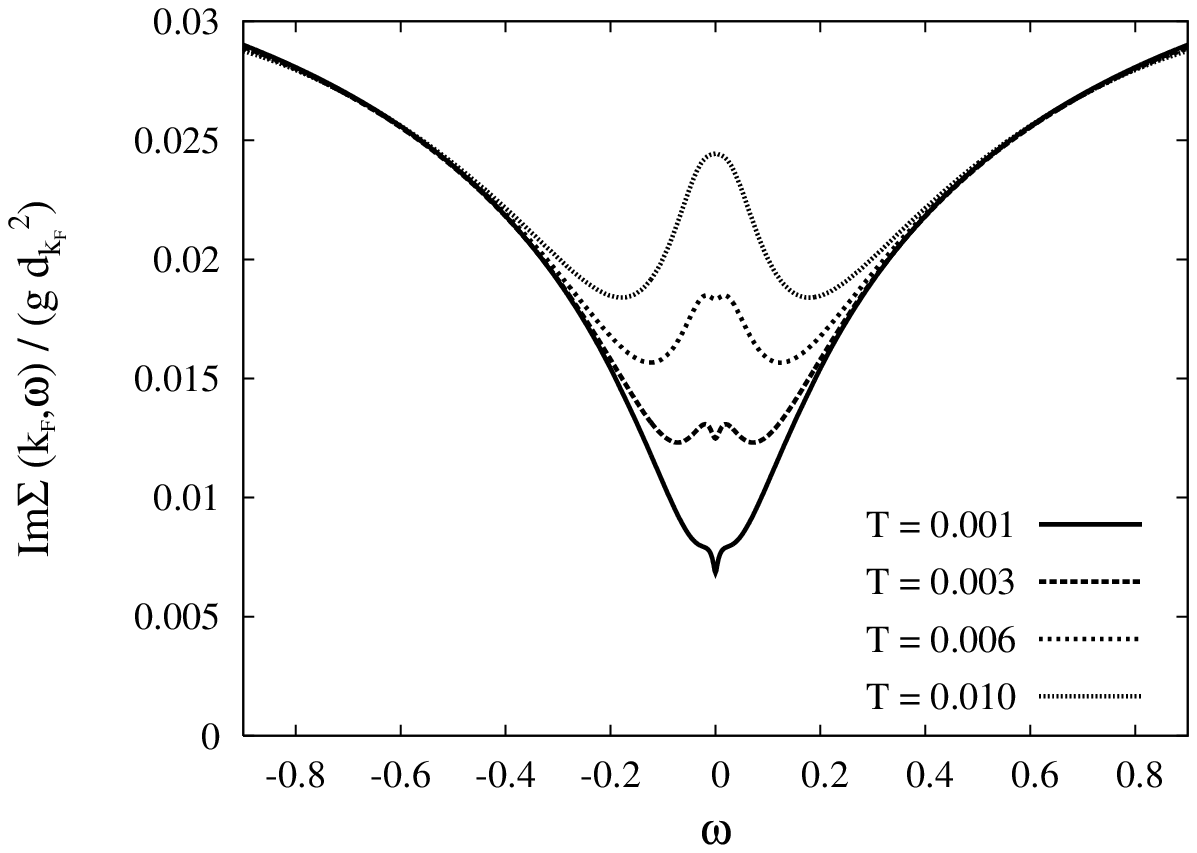,width=10cm}
\epsfig{file=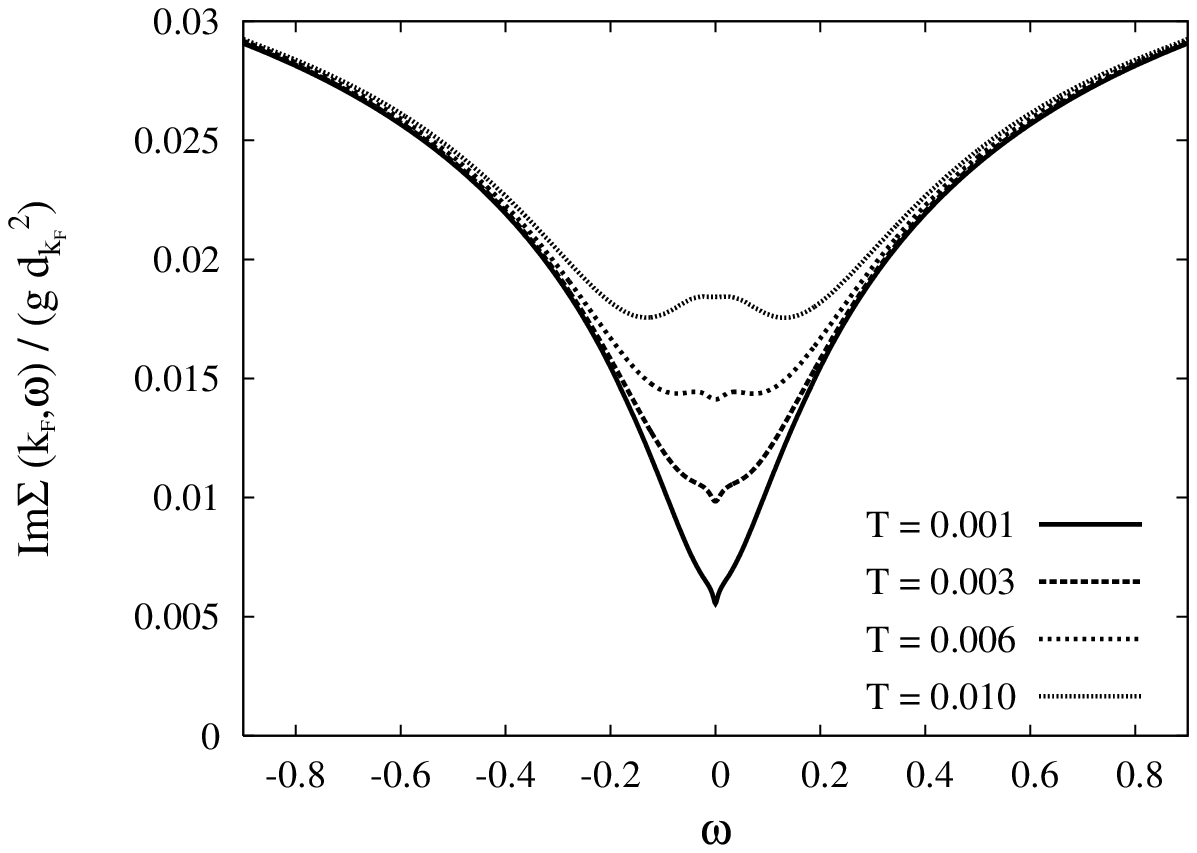,width=10cm}
\caption{Selfconsistent RPA results for the imaginary part of 
 the self-energy $\Im\Sg(\bk_F,\om)$, divided by $g d_{\bk_F}^2$, 
 as a function of $\om$ for several temperatures $T$, with a 
 correlation length $\xi(T) = 3/|T \log T|^{1/2}$ and
 $\xi_0 = v_{\bk_F} = u = 1$. Upper panel: $|g|d_{\bk_F}^2 = 1$,
 lower panel: $|g|d_{\bk_F}^2 = 4$.} 
\end{figure}

\begin{figure}
\center
\epsfig{file=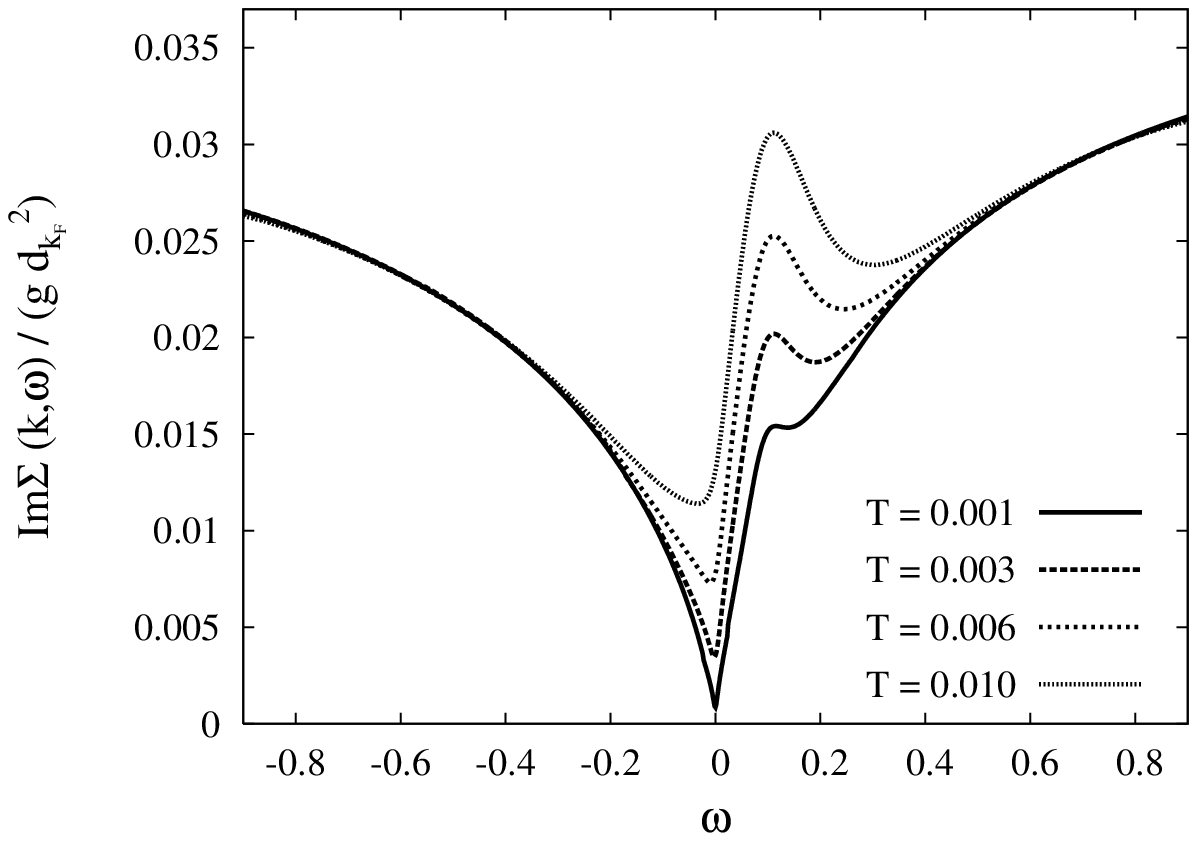,width=10cm}
\epsfig{file=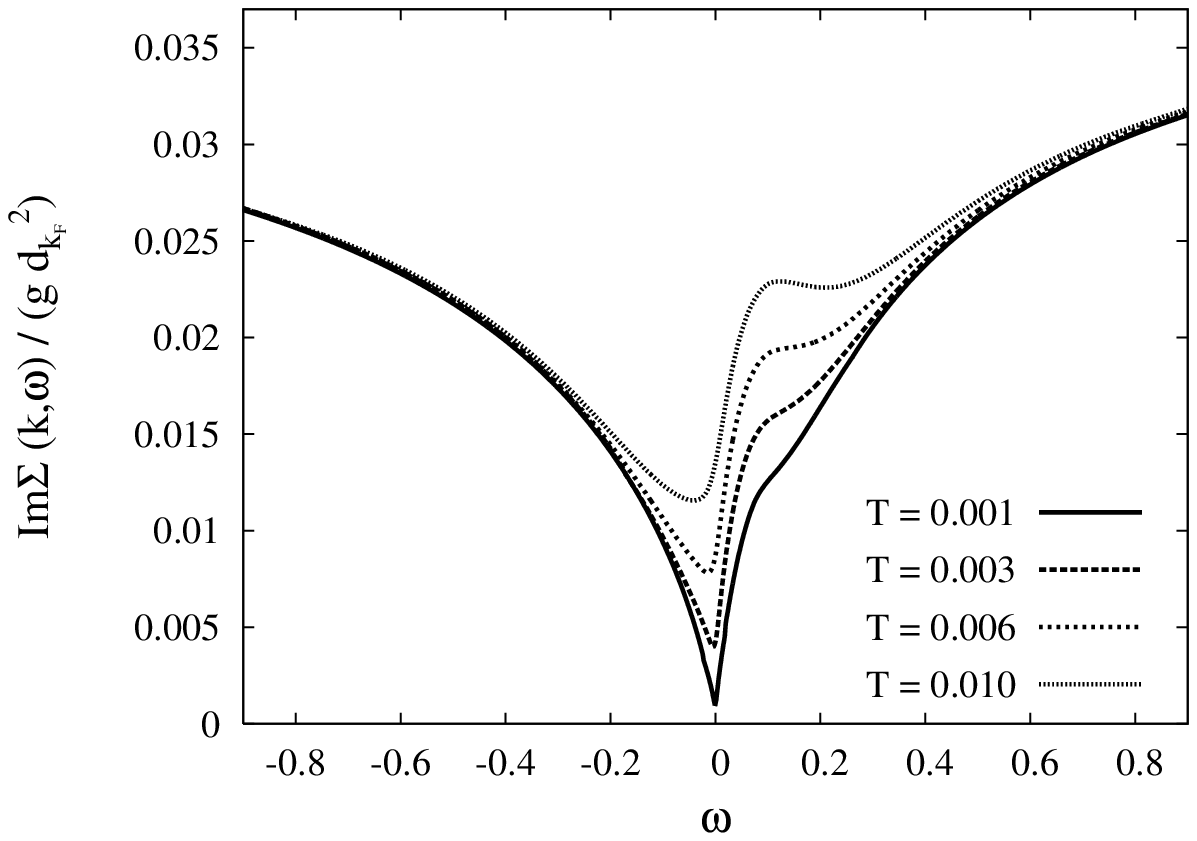,width=10cm}
\caption{Self-consistent RPA result for $\Im\Sg(\bk,\om)$ as
 in Fig.\ 8, but now for $k_r = 0.1$.} 
\end{figure}

\begin{figure}
\center
\epsfig{file=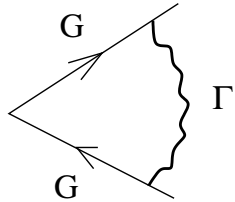,width=3cm}
\caption{Feynman diagram representing the first order vertex
 correction.} 
\end{figure}

\begin{figure}
\center
\epsfig{file=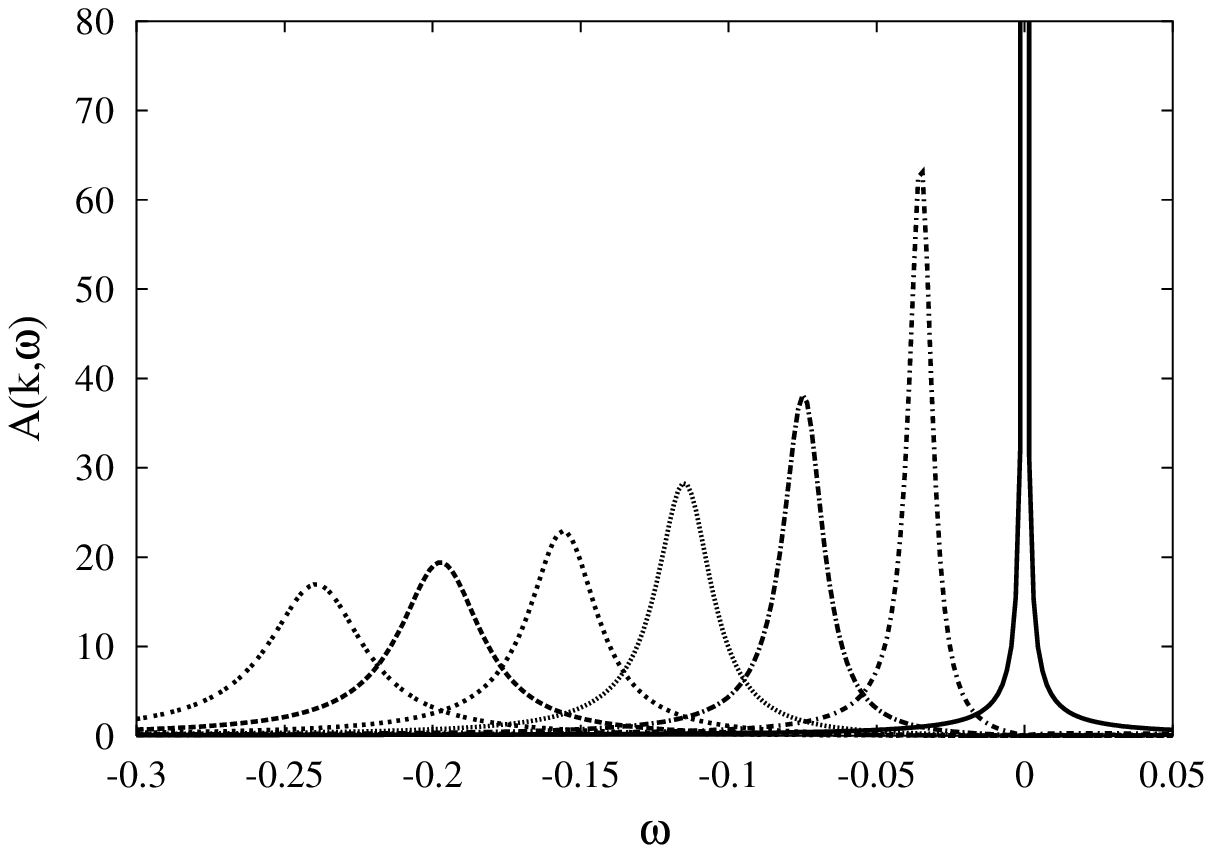,width=10cm}
\epsfig{file=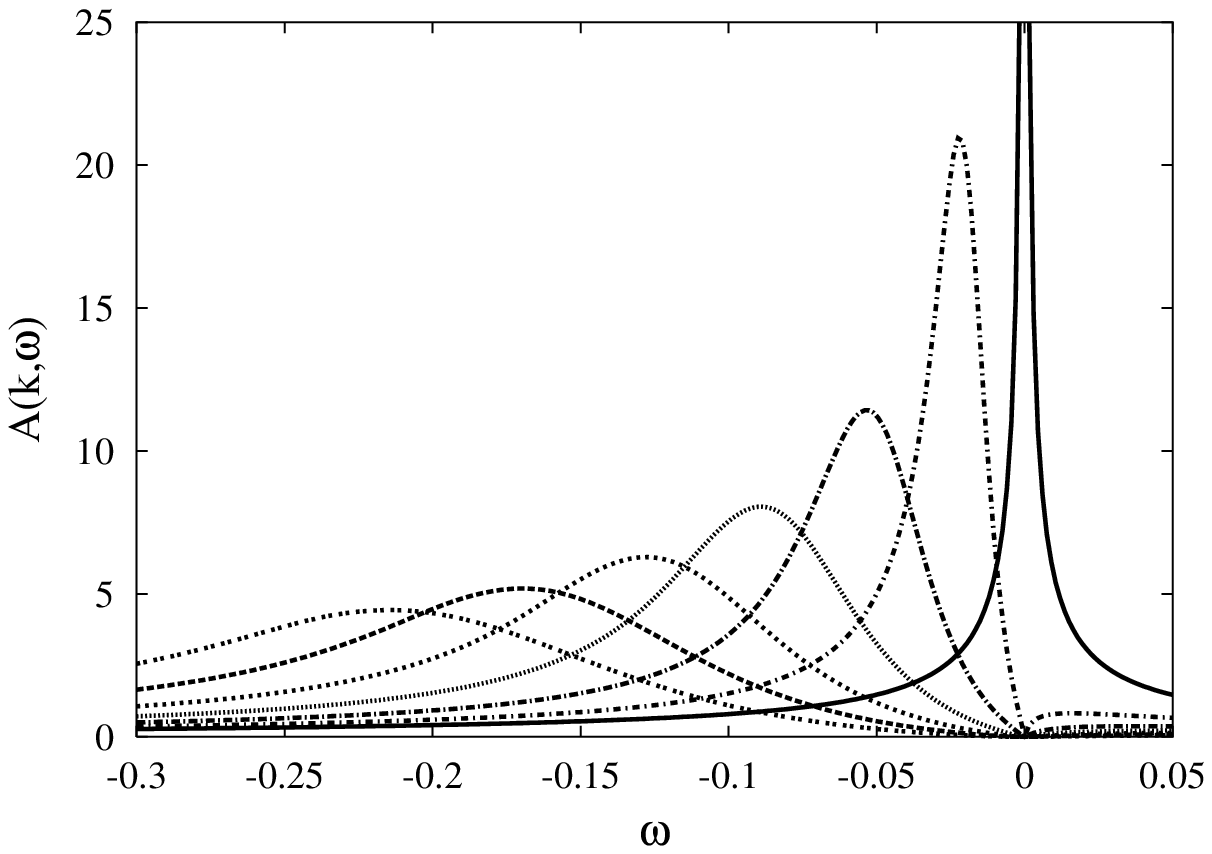,width=10cm}
\caption{Spectral function $A(\bk,\om)$ at the quantum critical 
 point as a function of $\om$ for
 $k_r = -0.0405 \, n$ with $n = 0,1,2,\dots,6$.
 Fixed parameters are $\xi_0 = v_{\bk_F} = u = 1$.
 Upper panel: $|g|d_{\bk_F}^2 = 1$, lower panel: 
 $|g|d_{\bk_F}^2 = 4$.} 
\end{figure}

\begin{figure}
\center
\epsfig{file=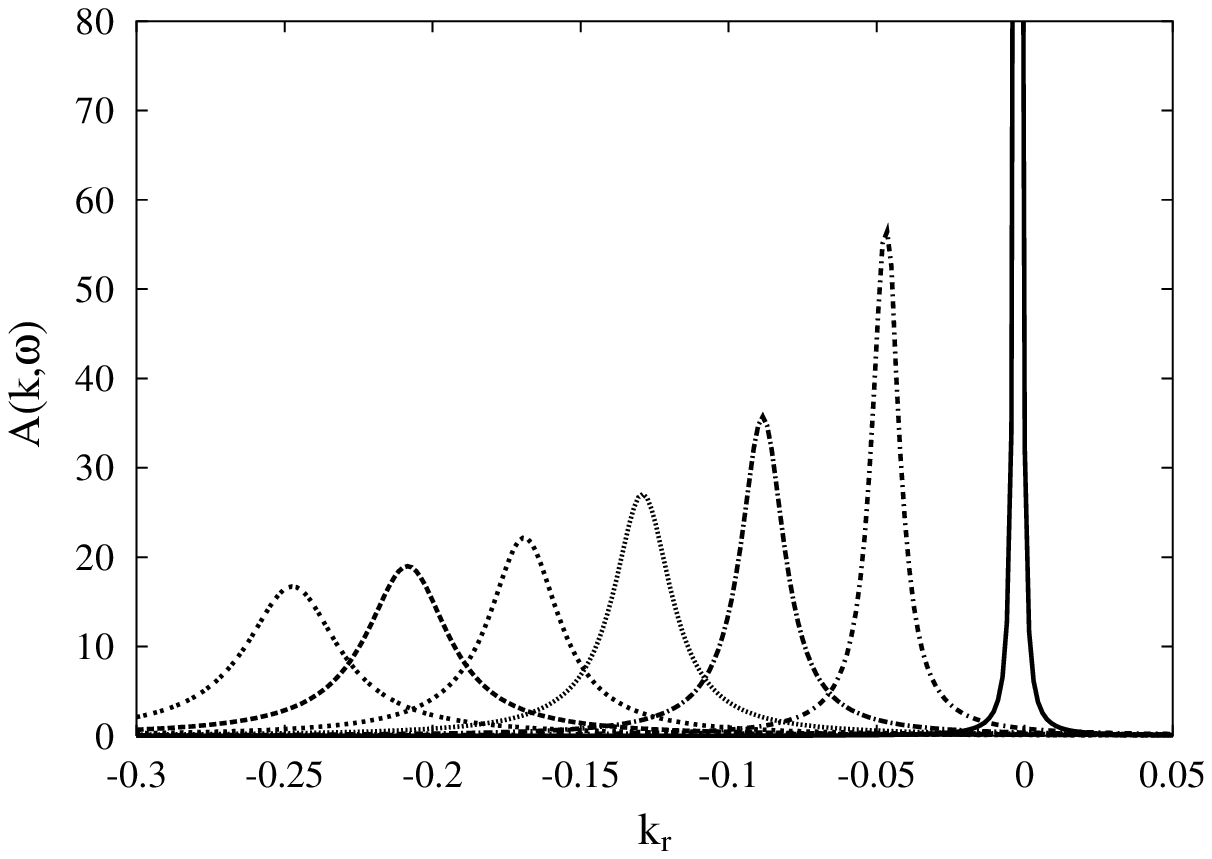,width=10cm}
\epsfig{file=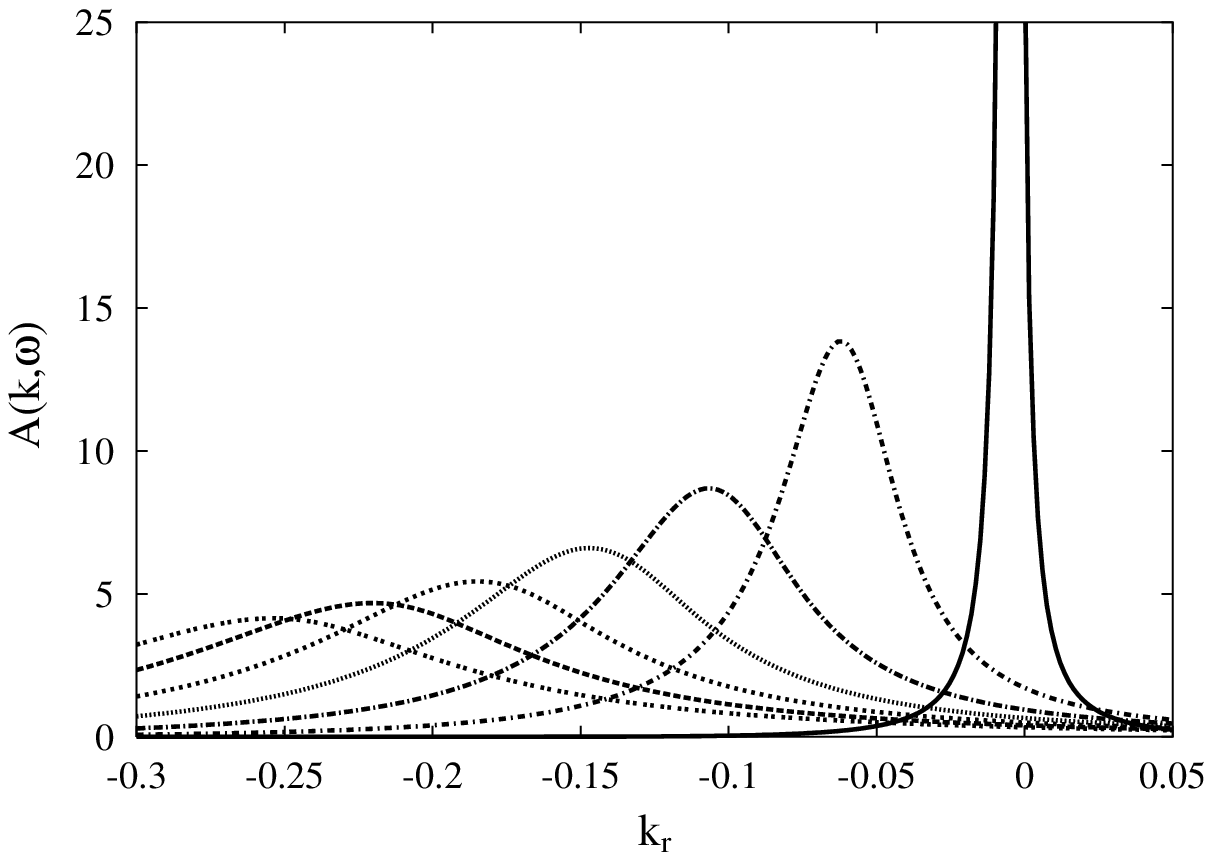,width=10cm}
\caption{Momentum scans of $A(\bk,\om)$ at the quantum critical 
 point for $\om = -0.0015 - 0.0405 \, n$ with $n = 0,1,2,\dots,6$. 
 The parameters are the same as in Fig.\ 11.} 
\end{figure}

\begin{figure}
\center
\epsfig{file=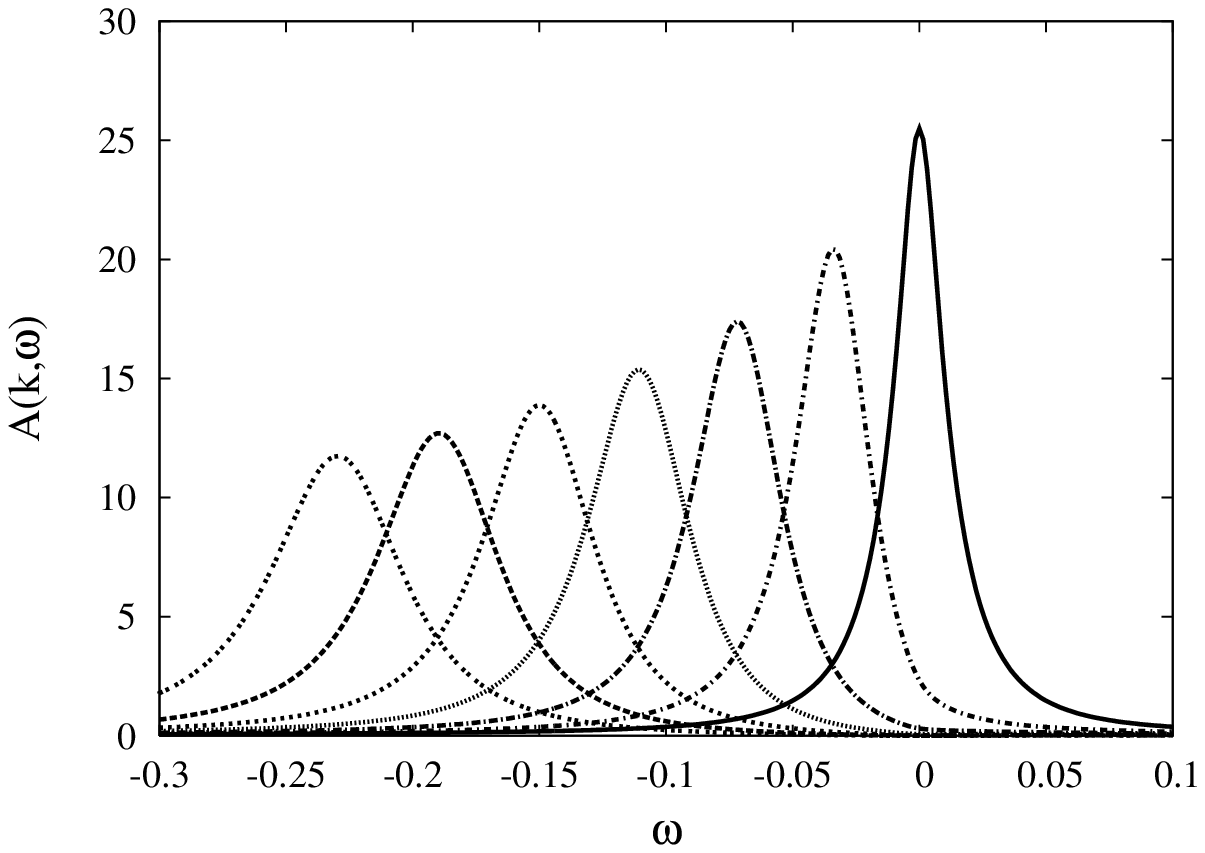,width=10cm}
\epsfig{file=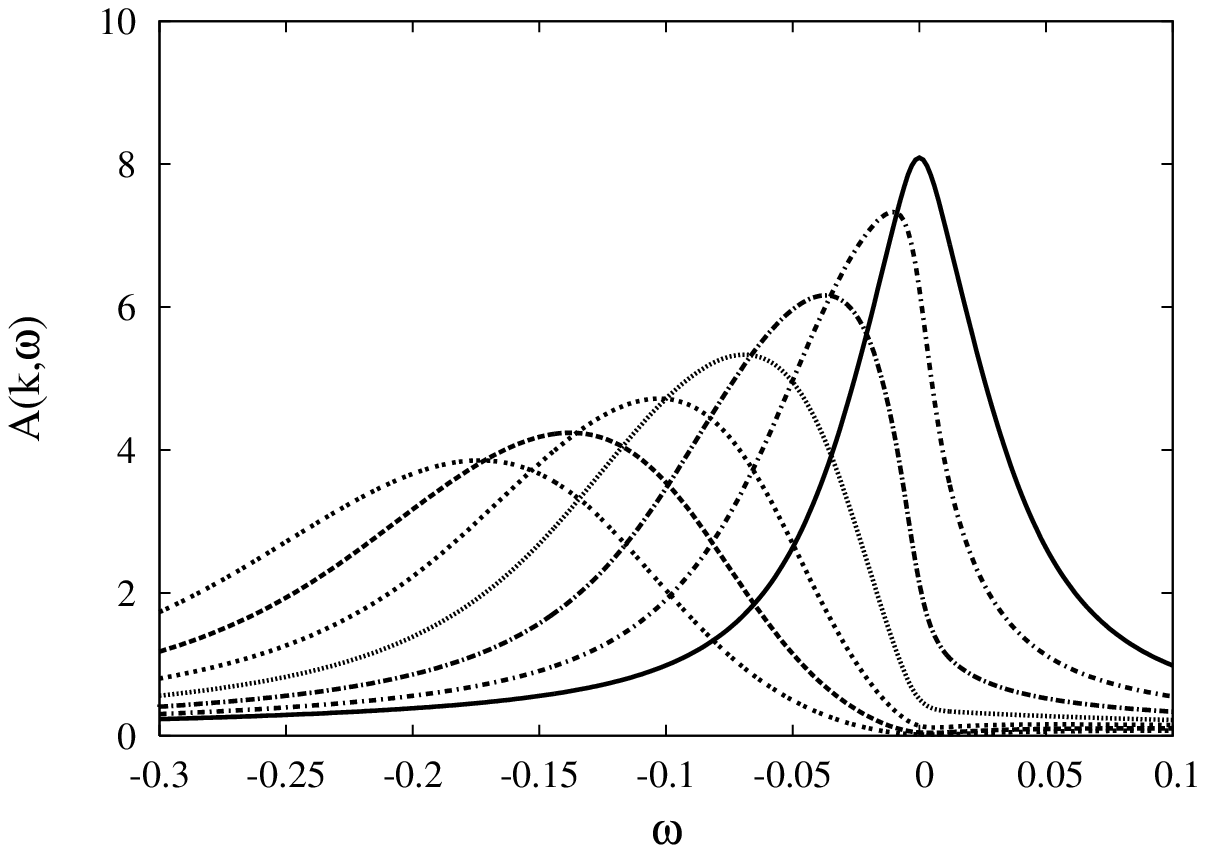,width=10cm}
\caption{Spectral function $A(\bk,\om)$ at $T = 0.003$ 
 as a function of $\om$ for 
 $k_r = -0.0405 \, n$ with $n = 0,1,2,\dots,6$.
 Parameters and $\xi(T)$ as in Fig.\ 8.} 
\end{figure}

\begin{figure}
\center
\epsfig{file=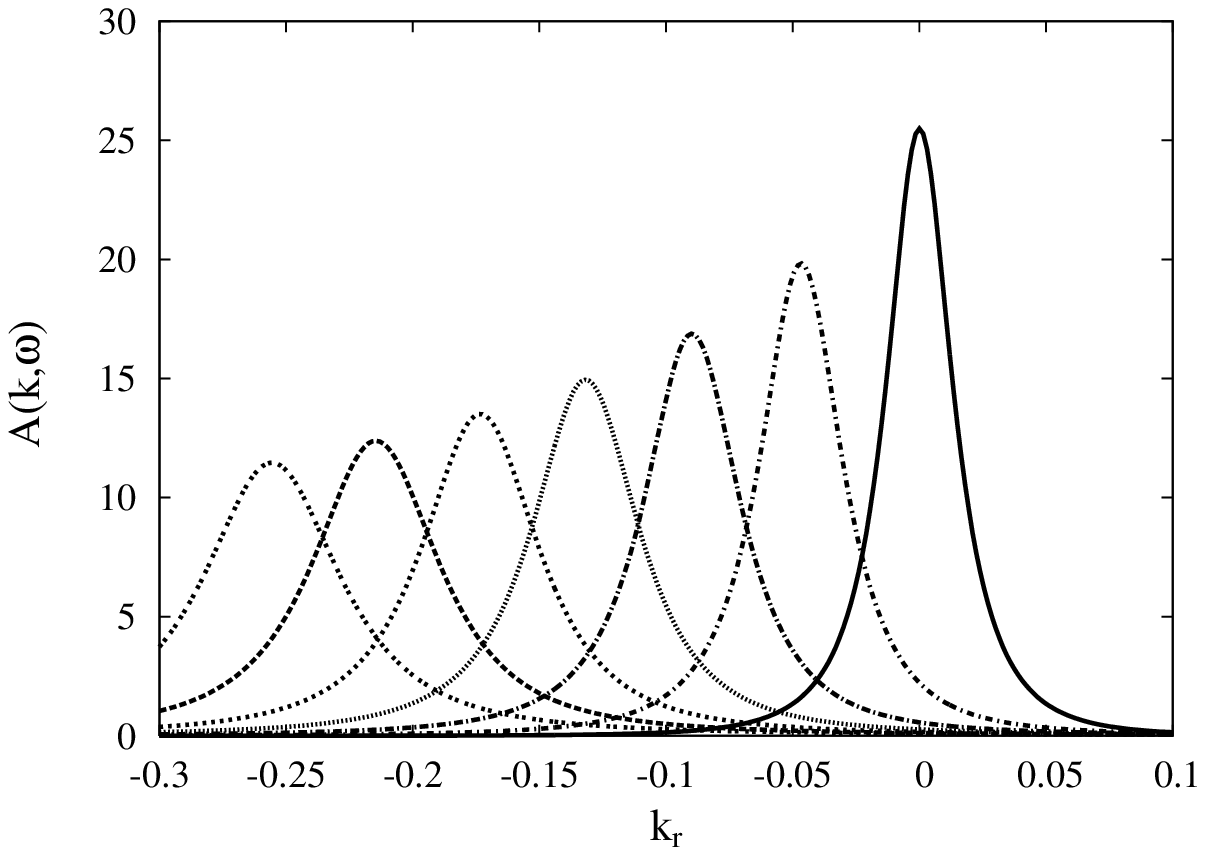,width=10cm}
\epsfig{file=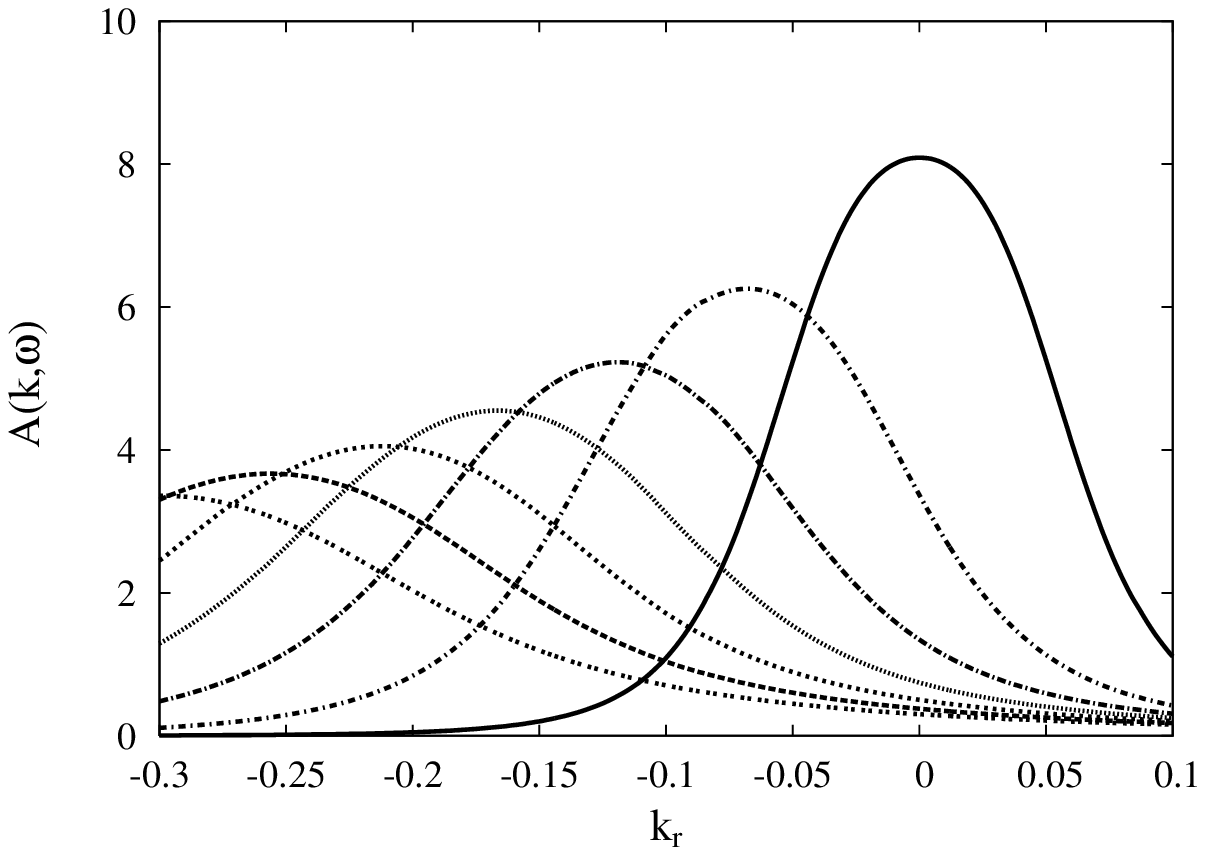,width=10cm}
\caption{Momentum scans of $A(\bk,\om)$ at $T=0.003$ for 
 $\om = - 0.0405 \, n$ with $n = 0,1,2,\dots,6$. 
 Parameters and $\xi(T)$ as in Fig.\ 8.} 
\end{figure}


\begin{thebibliography}{99}
\bibitem{fn1} The velocity $u(\hat\bq)$ shouldn't be confused
 with the interaction $u(\bq)$ introduced in Sec.~II.
\end{thebibliography}
\end{document}